\documentclass[twocolumn]{aastex61hack}

\newcommand{\rii}{\emph{r}-II}
\newcommand{\ri}{\emph{r}-I}
\newcommand{\ro}{\emph{r}-0}
\newcommand{\limr}{limited-\emph{r}}
\newcommand{\rp}{\emph{r}-process}

\renewcommand{\replaced}[2]{\textcolor{red}{%
\ifbib\let\sout\relax\fi\let\ref\specialref%
\let\citep\specialcitep%
\let\citet\specialcitet%
\let\cite\specialcite\sout{#1}
}\textcolor{blue}{#2}}

\usepackage{enumerate}
\usepackage{graphicx}
\usepackage{amsmath}
\usepackage{natbib}
\usepackage{float}
\usepackage{xcolor}
\usepackage{hyperref}
\hypersetup{
    bookmarks=true,                 
    unicode=false,                  
    pdftoolbar=true,                
    pdfmenubar=true,                
    pdffitwindow=true,              
    pdfstartview={FitH},            
    pdftitle={Reconstructing NSM Masses}, 
    pdfauthor={eholmbeck},          
    pdfsubject={Astrophysics},    
    pdfcreator={dvipdf},            
    pdfproducer={dvipdf},           
    pdfkeywords={r-process, stars}, 
    pdfnewwindow=true,              
    colorlinks=true,                
    linkcolor=magenta,              
    citecolor=teal,                 
    filecolor=magenta,              
    urlcolor=cyan,                  
    breaklinks=true,
    linktocpage
}

\begin{document}

\title{Reconstructing Masses of Merging Neutron Stars from Stellar \emph{R}-Process Abundance Signatures}

\author[0000-0002-5463-6800]{Erika M.\ Holmbeck}
\affiliation{Department of Physics, University of Notre Dame, Notre Dame, IN 46556, USA}
\affiliation{JINA Center for the Evolution of the Elements, USA}

\author[0000-0002-2139-7145]{Anna Frebel}
\affiliation{Department of Physics and Kavli Institute for Astrophysics and Space Research, Massachusetts Institute of Technology, Cambridge, MA 02139, USA}
\affiliation{JINA Center for the Evolution of the Elements, USA}

\author[0000-0001-6811-6657]{G.\ C.\ McLaughlin}
\affiliation{Department of Physics, North Carolina State University, Raleigh, NC 27695, USA}
\affiliation{JINA Center for the Evolution of the Elements, USA}

\author[0000-0002-4729-8823]{Rebecca Surman}
\affiliation{Department of Physics, University of Notre Dame, Notre Dame, IN 46556, USA}
\affiliation{JINA Center for the Evolution of the Elements, USA}

\author[0000-0003-4619-339X]{Rodrigo Fern\'{a}ndez}
\affiliation{Department of Physics, University of Alberta, Edmonton, AB T6G 2E1, Canada}

\author[0000-0002-4670-7509]{Brian D.\ Metzger}
\affiliation{Department of Physics, Columbia University, New York, NY 10027, USA}

\author[0000-0002-9950-9688]{Matthew R.\ Mumpower}
\affiliation{Theoretical Division, Los Alamos National Laboratory, Los Alamos, NM, 87545, USA}
\affiliation{Center for Theoretical Astrophysics, Los Alamos National Laboratory, Los Alamos, NM, 87545, USA}\affiliation{JINA Center for the Evolution of the Elements, USA}

\author[0000-0002-4375-4369]{Trevor M.\ Sprouse}
\affiliation{Theoretical Division, Los Alamos National Laboratory, Los Alamos, NM, 87545, USA}
\affiliation{Center for Theoretical Astrophysics, Los Alamos National Laboratory, Los Alamos, NM, 87545, USA}\affiliation{JINA Center for the Evolution of the Elements, USA}

\correspondingauthor{Erika M.\ Holmbeck}
\email{erika.holmbeck@rit.edu}


\begin{abstract}
Neutron star mergers (NSMs) are promising astrophysical sites for the rapid neutron-capture (``\emph{r}-") process, but can their integrated yields explain the majority of heavy-element material in the Galaxy?
One method to address this question has utilized a forward approach that propagates NSM rates and yields along with stellar formation rates, in the end comparing those results with observed chemical abundances of \rp-rich, metal-poor stars.
In this work, we take the inverse approach by utilizing \rp-element abundance ratios of metal-poor stars as input to reconstruct the properties---especially the masses---of the neutron star (NS) binary progenitors of the \rp\ stars.
This novel analysis provides an independent avenue for studying the population of the original neutron star binary systems that merged and produced the \rp\ material incorporated in Galactic metal-poor halo stars.
We use ratios of elements typically associated with the limited-\emph{r} process and the actinide region to those in the lanthanide region (i.e., Zr/Dy and  Th/Dy) to probe the NS masses of the progenitor merger. We find that NSMs can account for all \rp\ material in metal-poor stars that display \rp\ signatures, while simultaneously reproducing the present-day distribution of double-NS (DNS) systems.
However, the most \rp\ enhanced stars (the \rii\ stars) on their own would require progenitor NSMs of very asymmetric systems that are distinctly different from present ones in the Galaxy.
As this analysis is model-dependent, we also explore variations in line with future expectation regarding potential theoretical and observational updates, and comment on how these variations impact our results.
\end{abstract}

\keywords{nuclear reactions, nucleosynthesis, abundances -- stars: abundances -- stars: Population II -- binaries: close -- stars: neutron}

\received{\today}
\submitjournal{\apj}

\section{Introduction}

How were the heaviest, naturally occurring elements made?
From a nuclear physics perspective, this question has been answered for decades: through the rapid neutron-capture (``\emph{r}") process.
First introduced and named in the literature by \citet{b2fh} and \citet{cameron1957}, the \rp\ is thought to be one of the main mechanisms by which heavy, beyond-iron isotopes are synthesized.
The astrophysical question---where the \rp\ can naturally occur---is less obvious to address.

Ever since the \rp\ was identified as a major nucleosynthesis mechanism, 
neutron star mergers (NSMs) have been repeatedly proposed \citep{lattimer1974} and eventually confirmed \citep{abbott2017,cowperthwaite2017,drout2017,kilpatrick2017,shappee2017} as one source of \rp\ material in the universe.
What remains unclear is whether NSMs are sufficiently frequent or high-yield to account for the majority of \rp\ material in the Galaxy \citep{cote2018b}.

Additional observational evidence on the nature of \rp\ source(s) can be found in the abundance signatures of metal-poor stars.
Due to their chemical simplicity, metal-poor stars retain in their photospheres detectable imprints of individual nucleosynthetic events that occurred prior to their formation, at a time when the ISM had not yet been enriched by the fusion products of subsequent eras of star formation and stellar evolution \citep{barklem2005,beers2005,frebel2018}.
The elemental abundance patterns in metal-poor stars are therefore direct clues as to the lives and deaths of previous stellar generations and their remnants.
Stars with large amounts of \rp\ elements in their atmospheres are particularly helpful in this regard.
Many of these ``\rp-enhanced" stars indeed provide a nearly pure record of a single (to at most a few) prior astrophysical event that synthesized the ultimately observed heavy elements.

Investigations into potential, specific \rp\ conditions may manifest themselves in yield or relative abundance differences, that in turn, might be detectable within tangible star-to-star abundance variations. For example, several recent studies have investigated actinide production in NSMs in order to explain the subset of \rp-enhanced stars with high thorium and uranium abundances \citep{eichler2019,holmbeck2019b,holmbeck2019}.
By comparing theoretical \rp\ yields to observed stellar abundances, studies like these help to constrain the hydrodynamic and thermodynamic nature of the conditions under which the \rp\ occurs. Such constraints can then be used to explore particular sites for the \rp.

This study aims at taking such comparisons a step further by using observed abundances of \rp-enhanced stars together with results of recent simulations, to infer properties of the erstwhile NSs themselves. We assume that the \rp\ abundances in metal poor stars stem from single, prior mergers of two neutron stars, and that recent simulations are qualitatively good predictors of key quantities regarding element production. Yields of elements produced by the NSMs are then compared to the elemental signatures of the \rp-enhanced stars to calculate the masses of the progenitor double NS system (DNS) that merged and synthesized the heavy elements observed in the \rp-enhanced stars. We note that since the nuclear equation of state (EOS) is a known uncertainty in the outcome of hydrodynamic simulations, we include EOS effects in our investigation of whether NSMs (and which ones) were responsible for the majority of \rp\ production in the Galaxy. This innovative utilizing of observational abundances has the potential to forge new connections between the body of observed metal-poor stars, theoretical NSM studies, and upcoming results from the LIGO collaboration.

First, we discuss in Section~\ref{sec:properties} how NSM outflows are fundamentally related to the masses of the coalescing NSs and the nuclear EOS.
Section~\ref{sec:method} then describes how we will connect the merger outflow to stellar abundances of metal-poor stars in order to reconstruct the progenitor merging NS pair.
Section~\ref{sec:results} presents the results of this new method in terms of individual masses and mass distributions of the merging NS pair for six different EOSs, and the implications of these results are discussed in Section~\ref{sec:interpretations}.
Lastly, Section~\ref{sec:variations} explores variations on the model that can potentially support or oppose the theory that NSMs synthesized the majority of \rp\ material found in Galactic metal-poor stars.

\section{Puzzle Pieces: Neutron Star Properties}
\label{sec:properties}

Hydrodynamical simulations of NSMs predict the mass of the accretion disk around the merger remnant, the amount of dynamically ejected material, and the lifetime of the hypermassive neutron star (HMNS) before it collapses into a black hole (if at all).
These quantities bear directly on the 
global \rp\ abundances that may be ejected from individual NSMs and consequently enrich the star-forming ISM.
The increasing amount of NSM simulation data---and, with it, analytical descriptions of NSM ejecta---presents an opportunity to connect stellar \rp\ abundance signatures to progenitor NSM events.
With \rp\ abundances in hand from stellar signatures, we investigate the implication of assuming an NSM origin for the majority of \rp\ abundance in the Galaxy.
In the following sections, we review relevant parameterizations of hydrodynamical output and how we will use these analytic forms to find the individual masses of binary NS members.

\subsection{NSM Ejecta}
\label{sec:ejecta}

Material that may undergo nucleosynthesis and escape from an NSM is typically grouped into two main categories: the dynamical ejecta, which escapes promptly from the merger, and the wind outflows, which emanate from the accretion disk around the newly formed merger remnant.
Within the dynamical ejecta category, there exist multiple physical mechanisms that drive the ejection.
First, as the NSs coalesce, they become tidally deformed, and the tidal tails of the deformed star(s) are expelled from the system. 
Additionally, as the NSs make contact, a contact-interface ejecta can be produced.
There are also different physical mechanisms driving the wind outflows.
On longer timescales than the dynamical ejecta, an accretion disk/torus forms around the merger remnant, and material can be lost from the disk due to viscously heating, neutrino-driven winds, and magnetic stress.
A critical ingredient in our study is how much mass is ejected by each mechanism, which is largely determined by the masses of the colliding NSs.

\subsubsection{The Nuclear Equation of State}

Related to how NS masses shape NSM outflows is the nuclear EOS.
The EOS describes the behavior of ultra compact nuclear material and by extension determines the mass-radius relationship of neutron stars and the maximum neutron star mass.
The nuclear EOS is a subject of active investigation and a diverse array of efforts, theoretical, experimental and observational have been used to constrain it \citep[e.g.,][]{piekarewicz2009,lattimer2012,mathews2013,steiner2013b,radice2018}.
Table~\ref{tab:eos_params} lists the maximum non-rotating neutron star mass, as well as the radius of a neutron star at 1.4 and 1.6\,M$_\odot$ for six theoretical EOSs.
They are listed approximately from lowest compactness (largest radius; stiffest) to highest compactness (smallest radius; softest).

\begin{deluxetable}{llrrr}
\tablecaption{The maximum, non-rotating NS masses and radii at 1.4 and 1.6~M$_\odot$ for six different EOSs.\label{tab:eos_params}}
\tablehead{%
\colhead{EOS} & \colhead{Reference} & \colhead{$M_{\rm TOV}$} & \colhead{$R_{1.4}$} & \colhead{$R_{1.6}$} \\
\colhead{} & \colhead{} & \colhead{(M$_\odot$)} & \colhead{(km)} & \colhead{(km)} %
}
\startdata
  H4    & \citet{lackey2006} & 2.02 & 13.75 & 13.63 \\
  DD2   & \citet{typel2010} & 2.42 & 13.26 & 13.30 \\
  ALF2  & \citet{alford2005} & 2.06 & 13.16 & 13.19 \\
  LS220 & \citet{lattimer1991} & 2.05 & 12.66 & 12.48 \\
  MPA1  & \citet{muther1987} & 2.46 & 12.47 & 12.51 \\
  SFHo  & \citet{steiner2013} & 2.06 & 11.92 & 11.79 \\
\enddata
\end{deluxetable}

The EOS also sensitively affects the amount of NSM ejecta and its properties. For example, a stiff EOS generally allows more tidal deformability that leads to a greater dynamically ejected mass, and vice-versa for a soft EOS.
Additionally, the lifetime of the remnant massive object before it may collapse into a black hole also depends on the EOS; an HMNS with a stiff EOS generally resists collapse, leading to a longer-lived supra/hyper-massive NS remnant \citep[for details, see][among others]{dai1998,dai2006,hotokezaka2013,dietrich2015,radice2018,margalit2019,margalit2019b,radice2020}. 
Of the six equations of state in Table~\ref{tab:eos_params} two, H4 and SFHo, have been reported to be strongly disfavored based on an analysis of AT2017gfo kilonova.
Nevertheless, we keep them in our study to explore a large range of stiff to soft behavior and to test if our method similarly rules out these EOSs.

In the next sections, we discuss the models of the disk outflows and dynamical ejecta in detail and how these two quantities can be parameterized in terms of the NS masses and the EOS.
	
 	\begin{figure*}[t]
 	\centering
     \includegraphics[width=\columnwidth]{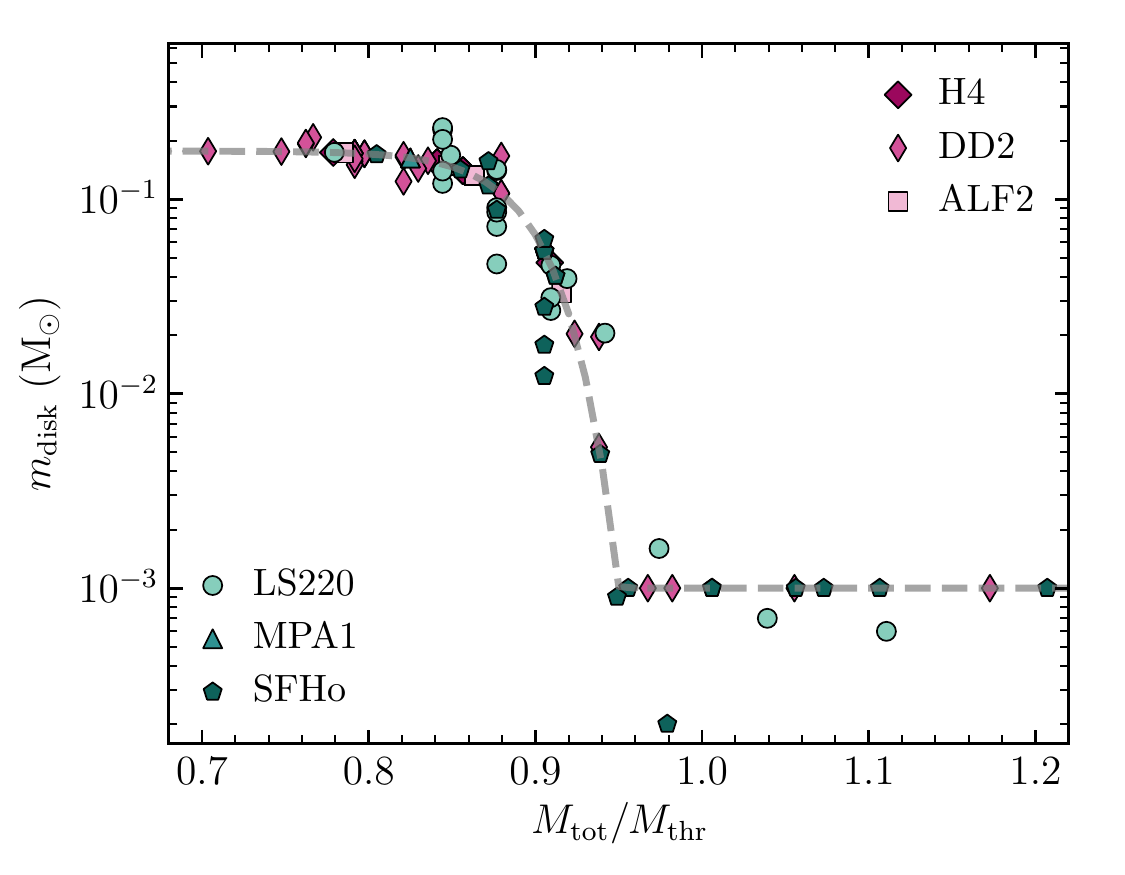}
     \includegraphics[width=\columnwidth]{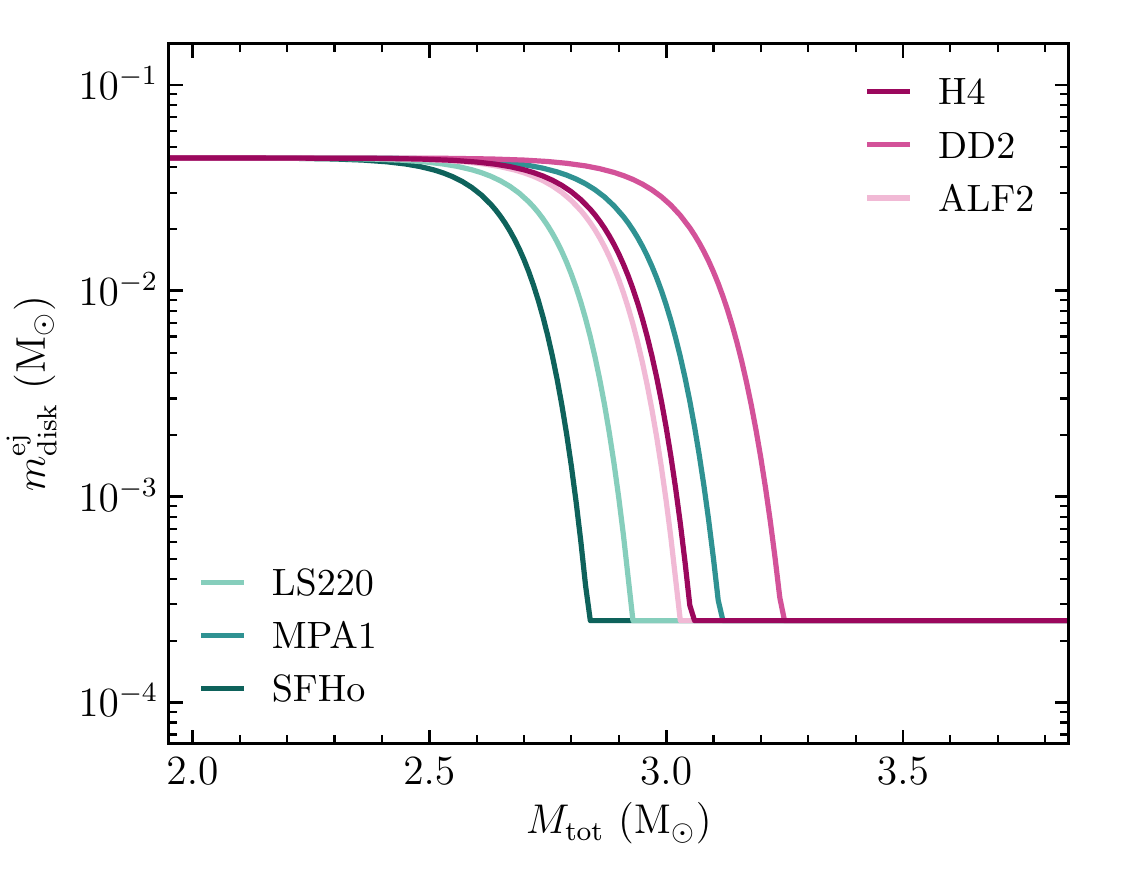}
     \caption{Analytic form for the mass of the accretion disk and the disk wind ejecta fit parameterized by Equation~\ref{eqn:disk} as a function of $M_{\rm tot}/M_{\rm thr}$ (left) and $M_{\rm tot}$ only (right). In the left panel, the gray, dashed curve shows this fit minimized over the EOS dependence. This EOS-minimized fit is compared to simulation data by the colored points found in \citet{dietrich2017} and \citet{radice2018} for different EOSs. The right panel displays the individual predicted disk wind ejecta for each of the six EOSs we use in this work.\label{fig:mdisk}}
 	\end{figure*}

\subsubsection{Disk Outflows}
	
The total mass lost through disk wind outflows depends on the mass of the accretion disk/torus that is created around the merger remnant.
Currently, only a few analytic fits of the disk mass to simulation data are available: Equation~25 in \citet{radice2018b}, Equation~1 in \citet{coughlin2019}, and Equation~4 in \citet{krueger2020}.
Both \citeauthor{radice2018b}\ and \citeauthor{coughlin2019}\ use the same set of numerical relativity simulation data, but the latter finds a lower fractional error between their fitting formula and simulation data than the fit by \citeauthor{radice2018b}.
The fit by \citeauthor{krueger2020} behaves similarly to the fit by \citeauthor{coughlin2019}, but we find that the latter produces slightly better agreement with simulation data for the EOSs considered in this work.
The fit from \citeauthor{coughlin2019}\ that we adopt approximates the disk mass as a function of the total mass, $M_{\rm tot} = M_1 + M_2$:
	\begin{equation}\label{eqn:disk}\begin{split}
	\log_{10}(m_{\rm disk}) = {\rm max}\bigg\{&-3, a\bigg(1+
	\\ & b\tanh \left[\frac{c-M_{\rm tot}/M_{\rm thr}}{d}\right]\bigg)\bigg\},
	\end{split}
	\end{equation}
with fitting parameters $a=-31.335$, $b=-0.9760$, $c=1.0474$, and $d=0.05957$, and $M_{\rm thr}$ given by
	\begin{equation}
	M_{\rm thr}=M_{\rm TOV}\left(2.38 - 3.606\frac{G M_{\rm TOV}}{c^2 R_{1.6\,\text{M}_\odot}}\right),
	\label{eqn:mthr}
	\end{equation}
where $M_{\rm TOV}$ is the maximum mass of a non-rotating NS (the Tolman-Oppenheimer-Volkoff limit), and $R_{1.6\,\text{M}_\odot}$ is the radius of a $1.6\,\text{M}_\odot$ star predicted by a given EOS.
The left panel of Figure~\ref{fig:mdisk}---which shows a plot similar to that in \citeauthor{coughlin2019}---shows Equation~\ref{eqn:disk} in gray compared to simulation data for many choices of EOS.
This figure reflects the general trend that slower collapse leads to a more massive accretion disk and therefore that less massive binary systems result in more massive disks. Using stiffer EOSs also typically results in a more massive disks, as can be seen in the right panel of Figure~\ref{fig:mdisk}.
These curves generally follow the order given in Table~\ref{tab:eos_params}, but that order is broken for EOSs with a large $M_{\rm TOV}$: DD2 and MPA1.
By Equation~\ref{eqn:disk}, a larger $M_{\rm TOV}$ allows more disk ejecta since these EOSs can support a more massive remnant.

In a merger, some fraction of the torus mass lost will be ejected. This fraction can be as little as a few percent or as high as 100\%, depending on the remnant lifetime \citet{metzger2014,lippuner2017,fernandez2018,coughlin2019}.
For disks surrounding a black hole, this ejecta fraction is rough 10--40\% \citep{siegel2018b,fujibayashi2020}.
For this initial study, we assume a constant of 25\% for the disk ejecta following \citet{just2015} and recognize this presents an area for improvement.
For easy reference, we have adjusted the curves in the right panel of Figure~\ref{fig:mdisk} by this 25\% and, for each EOS, we use the displayed estimate as the total mass of the disk outflows in the following sections.

\subsubsection{Dynamical Ejecta}

	\begin{figure*}[t]
	\centering
    \includegraphics[width=\columnwidth]{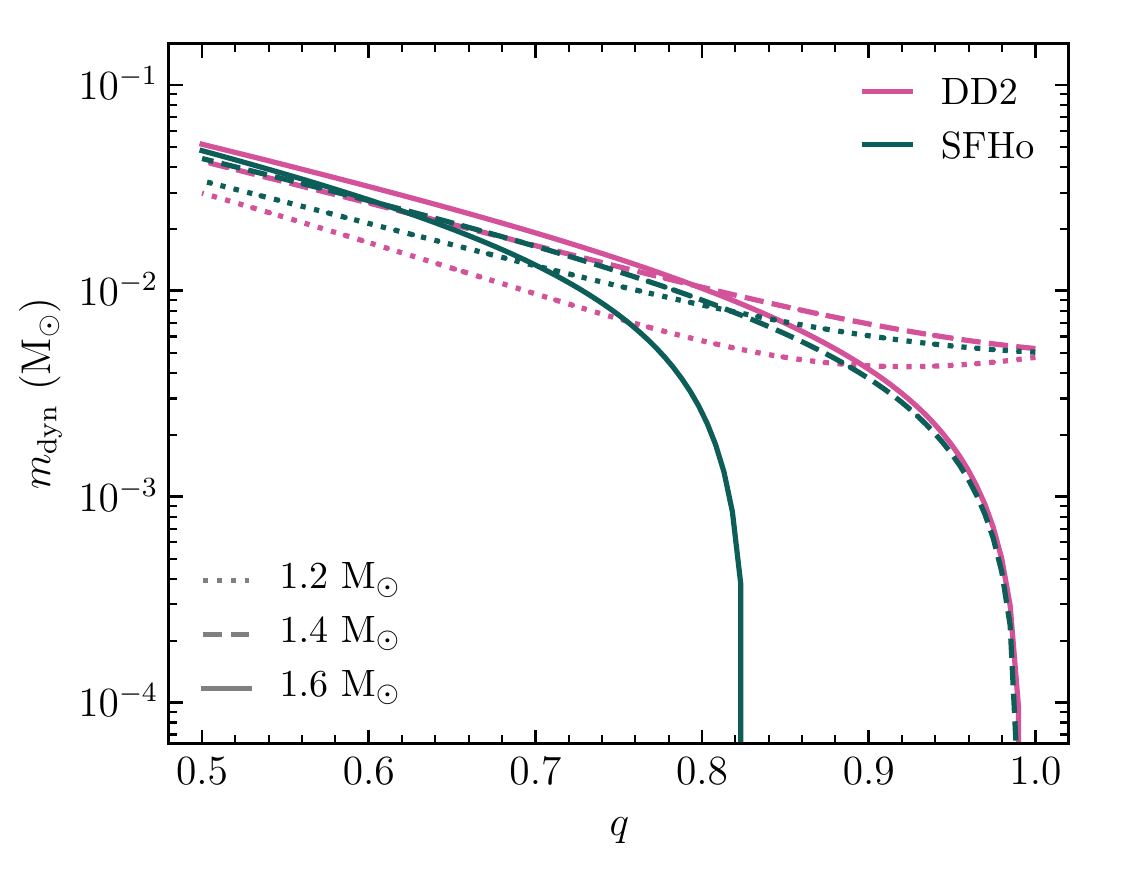}
    \includegraphics[width=\columnwidth]{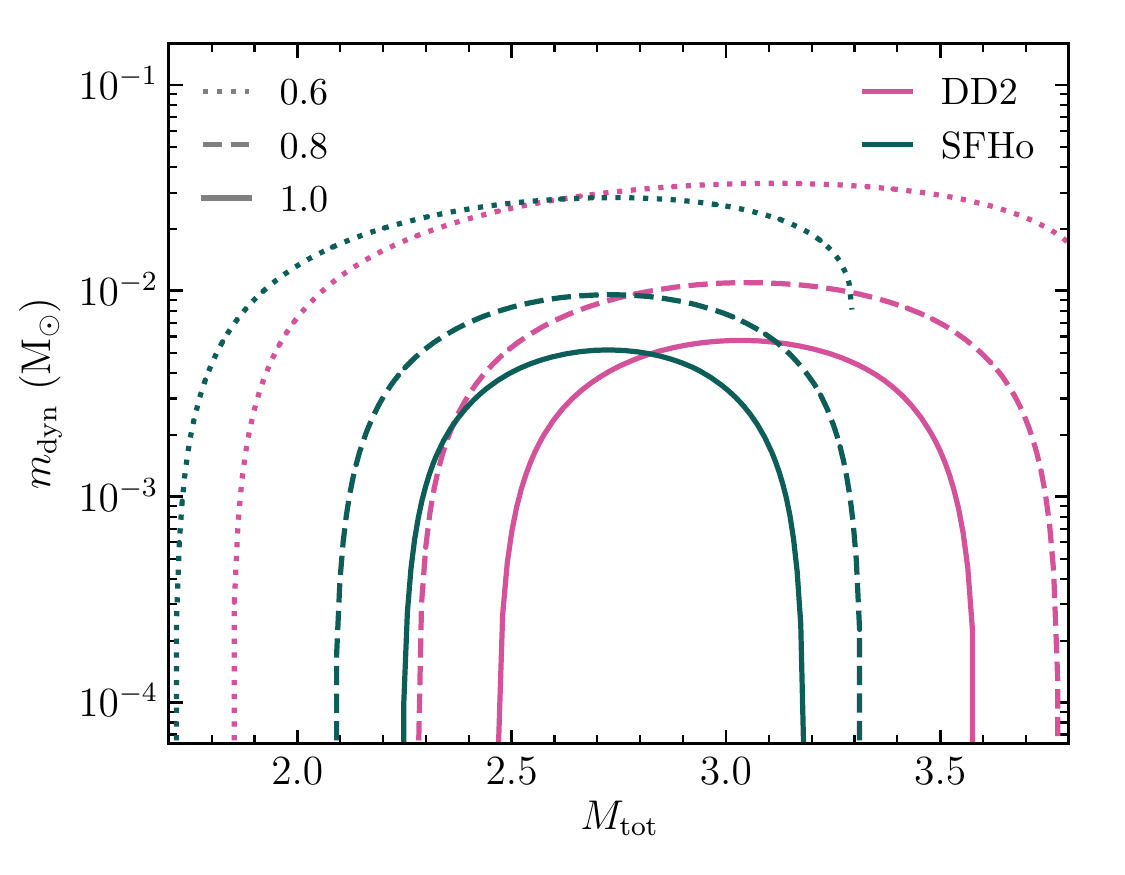}
	\caption{Analytic form for the mass of the dynamical ejecta (Equation~\ref{eqn:dynk}) comparing a stiff EOS (DD2, pink) and a soft EOS (MPA1, teal). The left panel displays this analytic form as a function of mass ratio, where each curve shows the calculation for a different value of the primary mass, $M_1$: 1.2\,M$_\odot$ (dotted), 1.4\,M$_\odot$ (dashed), and 1.6\,M$_\odot$ (solid). The right panel shows the predicted dynamical mass as a function of total mass applied to different mass ratios, $q$: 0.6 (dotted), 0.8 (dashed), and 1.0 (solid).\label{fig:mdyn}}
	\end{figure*}

We now discuss in more detail how the masses of the NSs and the EOS play a critical role in determining the dynamical ejecta, $m_{\rm dyn}$, focusing on the mass ratio in the binary system ($q\equiv M_2/M_1 \leq 1$), the total mass ($M_{\rm tot}$), and the EOS. There have been several fits to simulation data to determine the amount of dynamical ejecta: \citet{dietrich2017,radice2018b,coughlin2019,krueger2020}. Here we adopt the fit of \citet{krueger2020} as it was created for a wide range of masses, especially to accurately describe the behavior of dynamical ejecta at high NS masses:
	\begin{equation}
	\frac{m_{\rm dyn}}{10^{-3}\text{M}_\odot} = \left[\frac{a}{\mathcal C_1} + b\left(\frac{M_2}{M_1}\right)^n + c\mathcal C_1\right]M_1 + \left[1\leftrightarrow 2\right],
	\label{eqn:dynk}
	\end{equation}
where $M_i$ is the gravitational mass of the NS, and $\mathcal C_i$ is the compactness parameter defined as
	\begin{equation}
	\mathcal C_i = \frac{G M_i}{c^2R_i},
	\end{equation}
where $R_i$ is the radius of the NS at a mass $M_i$, which can be obtained from the mass-radius relationship solved for each EOS.
With this formulation, \citet{krueger2020} find best-fit values to simulation data of $a=-9.3335$, $b=114.17$, $c=-337.56$, and $n=1.5465$. Figure~\ref{fig:mdyn} shows this analytic form both as a function of the mass ratio $q$ (left) and total binary mass $M_{\rm tot}$ (right) for a relatively stiff (DD2) and very soft (SFHo) EOS.

First let us compare different curves of the same color (i.e., one EOS).
From the left panel Figure~\ref{fig:mdyn}, we see that generally a more extreme mass ratio ($q<1$) leads to more tidal deformation and hence more dynamical ejecta mass. For example, the range of maximum ejecta mass from $q=0.5$ to $q=1.0$ is roughly an order of magnitude and can drop precipitously to zero at higher NS masses.
This difference is also shown in the right panel of Figure~\ref{fig:mdyn}. 
Reasonably long plateaus exist for the dynamical ejected mass as a function of total mass, and they can be effectively shifted by about an order of magnitude by changing the mass ratio.
These plateaus also indicate that the total mass has little sensitivity on the dynamical ejecta, except at extreme values of total mass.
The steep drop to the left of the plateaus (low $M_{\rm tot}$) occur because systems with low total mass do not cause enough tidal deformability by which to dynamically eject mass. Remember, however, that a binary with the same total mass but a higher mass ratio will eject more mass. The rapid decline to the right of the plateaus are caused by very massive systems that are expected to collapse into a black hole more quickly than their less massive counterparts, allowing little time to for material to be ejected dynamically.
These effects also explain the steeps drops for high and low-$q$ in the left panel for the low and mass-mass curves, respectively.

Now we turn to the EOS effects---which enter here through the compactness parameter---on the dynamical ejecta mass by comparing the pink (DD2) and teal (SFHo) curves in Figure~\ref{fig:mdyn}.
These curves represent EOSs with comparatively low and high $\mathcal C$, respectively.
First compare dashed pink and teal lines in the left panel of Figure~\ref{fig:mdyn}.
Both curves represent binary NS systems with the same primary mass, $M_1=1.6$\,M$_\odot$, and the mass of the secondary determined by $M_2=1.6\,q$ (M$_\odot$).
For values of $q\gtrsim 0.7$, the dashed teal curve (soft; high $\mathcal C$) predicts a lower dynamical mass than the corresponding pink curve (stiff; low $\mathcal C$).
For even higher NS masses ($M_1=1.8$\,M$_\odot$), the soft EOS predicts no dynamical ejecta at all for $q\gtrsim 0.82$.

Now consider the plateaus in the right panel of Figure~\ref{fig:mdyn} again, comparing similar curves with different colors.
Notice now that plateaus with the same $q$ for different EOSs may overlap with an apparent shift to $M_{\rm tot}$.
In other words, assuming a stiff EOS for systems with, e.g, $q=1.0$ will eject roughly the same amount of dynamical material as a system with a smaller total mass and a higher compactness (softer EOS).
Therefore, binary NSs with higher compactness tend to eject less dynamical material than binaries of smaller compactness.
In general, note that the dynamically ejected mass stays relatively constant for a wide range of $M_{\rm tot}$.
By comparing the left and right panels of Figure~\ref{fig:mdyn}, we can see directly that it is primarily $q$ governing the mass dynamically ejected by an NSM.

This delicate interplay between the NS masses, their mass ratio, and EOS determines the amount of \rp\ material that may be ejected into the ISM and make its way into the primordial gas that formed metal-poor stars to eventually be found in observations of stellar spectra.
Before we implement these equations to theoretical NSMs however, we can first apply these outflow estimates to GW170817 to test if they agree with kilonova observables.

\subsection{Application to GW170817}

In this section, we apply Equations~\ref{eqn:disk} and \ref{eqn:dynk} to GW170817 using observationally derived constraints on $M_1$ and $M_2$. 
We also compare those mass ejecta estimates to similar quantities derived from the associated kilonova to verify our use of these equations.

\begin{figure}[t]
	\centering
    \includegraphics[width=\columnwidth]{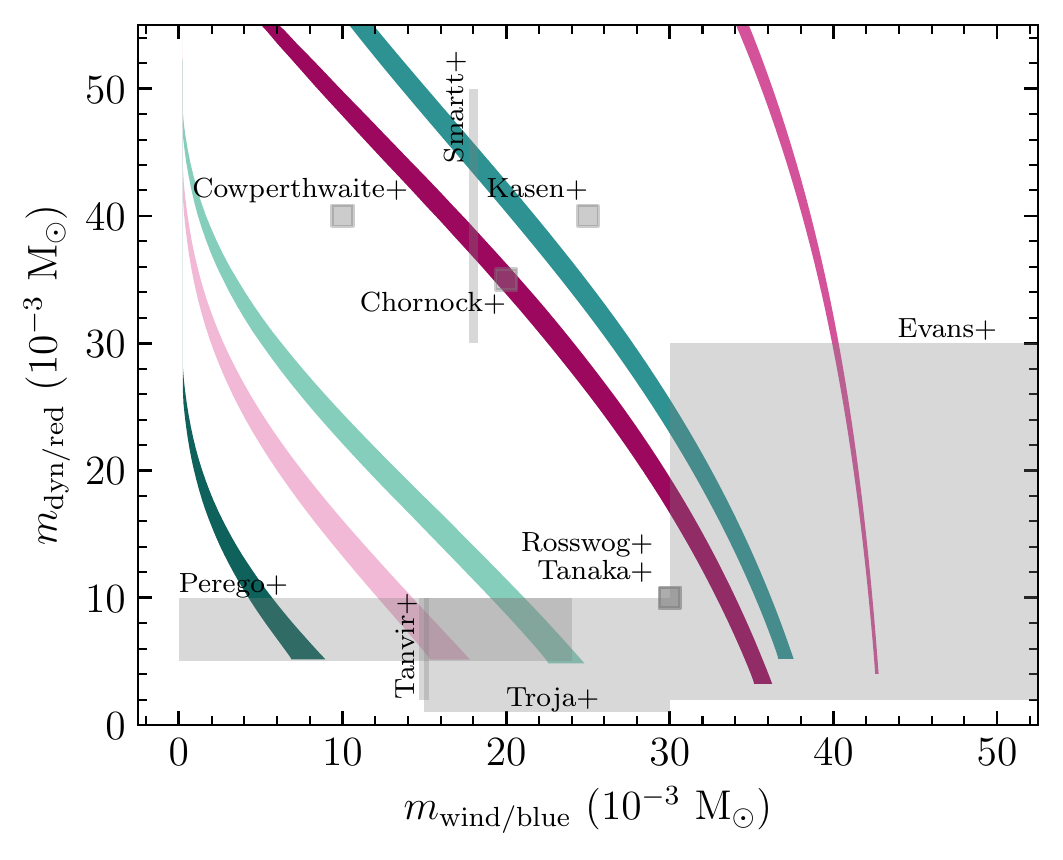}
    \caption[Predicted red or dynamical and blue or wind ejecta masses from GW170817 compared to literature values]{Predicted dynamical (or red) and wind (or blue) ejecta masses from GW170817 for each EOS considered in this work (colored curves, from left to right: SFHo, LS220, ALF2, H4, MPA1, and DD2) compared to literature values (gray boxes).\label{fig:gw170817}}
	\end{figure}

Equations~\ref{eqn:disk} and \ref{eqn:dynk} are structured as functions of the individual NS masses.
While precise estimates on the individual NS masses cannot be attained from present GW data for GW170817, the chirp mass offers a tight constraint on the {\it combination} of $M_1$ and $M_2$.
Therefore, we take all $M_1$-$M_2$ combinations that satisfy the chirp mass of GW170817 \citep[$\mathcal M=1.188_{-0.002}^{+0.004}$\,M$_{\odot}$;][]{abbott2017} and apply Equations~\ref{eqn:disk} and \ref{eqn:dynk} to obtain the wind and dynamical ejecta masses. In Figure~\ref{fig:gw170817} we represent all possible wind and dynamical mass combinations as colored bands from an NSM with GW170817's chirp mass for each of the six EOSs we use in this work.

We compare these curves to the various direct \citep{evans2017,perego2017,rosswog2017,tanaka2017,tanvir2017,troja2017} and indirect \citep{cowperthwaite2017,chornock2017,kasen2017,smartt2017} estimates of the wind and dynamical ejected masses that are derived from observations of the associated kilonova. These estimates are shown as gray boxes in Figure~\ref{fig:gw170817}.
Studies that place direct limits on the wind and dynamical ejecta masses fall primarily on the lower right of the plot.
Other studies of the kilonova instead infer the masses of {\it red} and {\it blue} kilonova components. 
We present these red and blue results on the same axes as the dynamical and wind masses, respectively. However, the mapping between red and blue to dynamical and wind is not so straightforward and hinges upon the details of the ejecta.
For example, depending on the $Y_e$, the wind can be red, blue, or ``purple" \citep[showing both red and blue components; see, e.g.,][]{villar2017}. Nevertheless, we can obtain a rough comparison with these studies by using a direct---though perhaps unrealistic---one-to-one mapping of the blue component to the wind outflow mass and the red component to the dynamical ejecta mass as in, e.g., \citet{Cote+18}. These estimates lie on the upper left of the plot.

Even though large uncertainties exist between literature studies of GW170817, systematic uncertainties due to the EOS also cause the theoretical estimates to predict very different mass yields.
Yet these theoretical curves cover roughly the same mass ranges as the estimates derived independently from the kilonova signal.
This agreement, while providing little constraint on the GW170817 precursor binary and EOS properties, at least verifies that Equations~\ref{eqn:disk} and \ref{eqn:dynk} reasonably reproduce values derived from an NSM observation.

The use of these equations is also validated by the second NSM to be observed by LIGO/Virgo: GW190425 \citep{abbott2020}.
This system was found to have a very high total mass, $M_{\rm tot}=3.4_{-0.1}^{+0.3}$ M$_{\odot}$; however, no electromagnetic afterglow was identified \citep{coughlin2019b,foley2020,pozanenko2020}.
At such a high total mass, Equations~\ref{eqn:disk} and \ref{eqn:dynk} both fall to minimum values.
In other words, very little to no luminous ejecta could be expected from such a high-mass merger.
Next, we turn to the composition of the outflowing ejecta and how the NS masses may effect the extent of the \rp\ pattern in addition to the abundance of \rp\ elements.

\subsection{Remnant Collapse Time}
\label{sec:lifetime}

In the previous sections, we have summarized the dependence of ejecta mass on the initial NS masses and the EOS. 
Not only do these quantities directly affect the amount of mass ejected, but they also determine the nuclear composition of the wind ejecta, which we outline here.
\citet{lippuner2017} showed that the time it takes for a HMNS remnant produced from an NSM to collapse into a black hole (its ``lifetime") affects the nucleosynthesis in the accretion disk outflows from the merger.
A longer collapse time leads to a less robust \rp\ pattern, i.e., lower relative lanthanide and actinide abundances.
The remnant lifetime affects the nucleosynthesis in this way due to the extent of allowed neutrino interactions that alter the composition of the ejecta \citep[e.g.,][]{metzger2014,martin2015,miller2019}.
The lifetime of the remnant before its collapse into a black hole sets a neutrino irradiation time, which can drive down the initial neutron-richness of the \rp\ composition (i.e., a higher $Y_e$), thus producing a less robust \rp\ pattern.

\begin{figure}[t]
	\centering
    \includegraphics[width=\columnwidth]{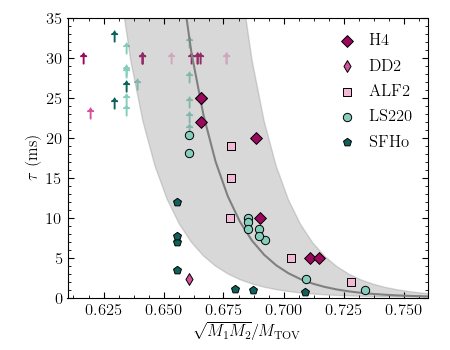}
    \caption{Lifetime of the HMNS remnant before it collapses into a black hole (if at all) as a function of $\sqrt{M_1 M_2}/M_{\rm TOV}$. Up arrows indicate the HMNS survived longer than the simulation time.\label{fig:tau_fit}}
	\end{figure}

The collapse time of the merger remnant also depends on the EOS and the total binary mass, as investigated in \citet{hotokezaka2013}.
\citet{lucca2019} compiled many investigations to fit a simple analytical form describing the collapse time from the mass and radius of the HMNS remnant.
	\begin{equation}
	\log_{10}(\tau) = e_0 + e_1 \log_{10} \left(\frac{\sqrt{M_1 M_2}}{M_{\rm TOV}}\right),
	\label{eqn:tau}
	\end{equation}
with best-fit parameters $e_0=-5.51\pm 0.36$ and $e_1=-39.0\pm 1.6$.
Recall that $M_{\rm TOV}$ is EOS-dependent.
Figure~\ref{fig:tau_fit} shows a variety of simulation data (from \citealt{hotokezaka2013}, \citealt{dietrich2015}, \citealt{kastaun2015}, and \citealt{radice2018}) plotted against Equation~\ref{eqn:tau}.
Lower limits are displayed as up-arrows and typically indicate that the lifetime exceeded the computation time in the simulation due to the difficulty of modeling such a complex system \citep[see, e.g.,][]{kiuchi2018}.
Computational limitations place significant uncertainties on the HMNS lifetime.

With Equations~\ref{eqn:disk}, \ref{eqn:dynk}, and \ref{eqn:tau}, it is now possible to estimate the total mass ejected from the dynamical mechanism and disk outflows as well as the ejecta composition using the masses of the NSs and a given EOS.
In summary, the total mass, mass ratio between the NSs, and the EOS affect the amount of ejected mass and remnant lifetime as follows:
	\begin{itemize}
	\item $M_{\rm tot}$: the mass of the accretion disk and the remnant lifetime are generally anti-correlated with the total mass, and the correlation with $m_{\rm dyn}$ is mostly flat except for very high or very low total masses (see Figure~\ref{fig:mdyn}).
	\item $q$: under Equations~\ref{eqn:disk} and \ref{eqn:tau}, the disk mass and remnant lifetime have no clear dependence on the mass ratio; however, the dynamical mass is generally anti-correlated with $q$, except for very high masses.
	\item EOS: both the ejecta masses as well as the remnant lifetime are anti-correlated with compactness, or, equivalently, positively correlated with tidal deformability.
	\end{itemize}
Next, we review the method by which we place estimates on the progenitor binary NS systems that merged to create the observed \rp\ enhancement of metal-poor stars in the Galactic halo.

\section{Method: Assembling the Puzzle}
\label{sec:method}

As described in Section~\ref{sec:ejecta}, a choice of NS masses and EOS will affect the amount of mass ejected by the merger.
In addition, the NS masses also determine the \rp\ extent of the ejecta, in particular of the disk wind outflows, by setting the degree of a neutrino irraditation from the HMNS lifetime.
The mass of ejected disk and dynamical material then sets the amount by which to scale the respective abundance patterns of the ejected material.
Therefore, given a choice of (independent variables) $M_1$, $M_2$, and EOS, we can quantify the total \rp\ ejecta from a merger event.

Here we describe this concept in reverse: how to go from observed \rp\ abundances to the original NS masses of the progenitor merger.
First, we describe the nucleosynthesis simulations used for the disk and dynamical ejecta.
Next, we detail the MCMC method we use to explore the $M_1$-$M_2$ parameter space to build a bridge between observationally derived \rp\ abundances and NSM properties.

\subsection{Nucleosynthesis Calculations}

\begin{figure}[t]
	\centering
    \includegraphics[width=\columnwidth]{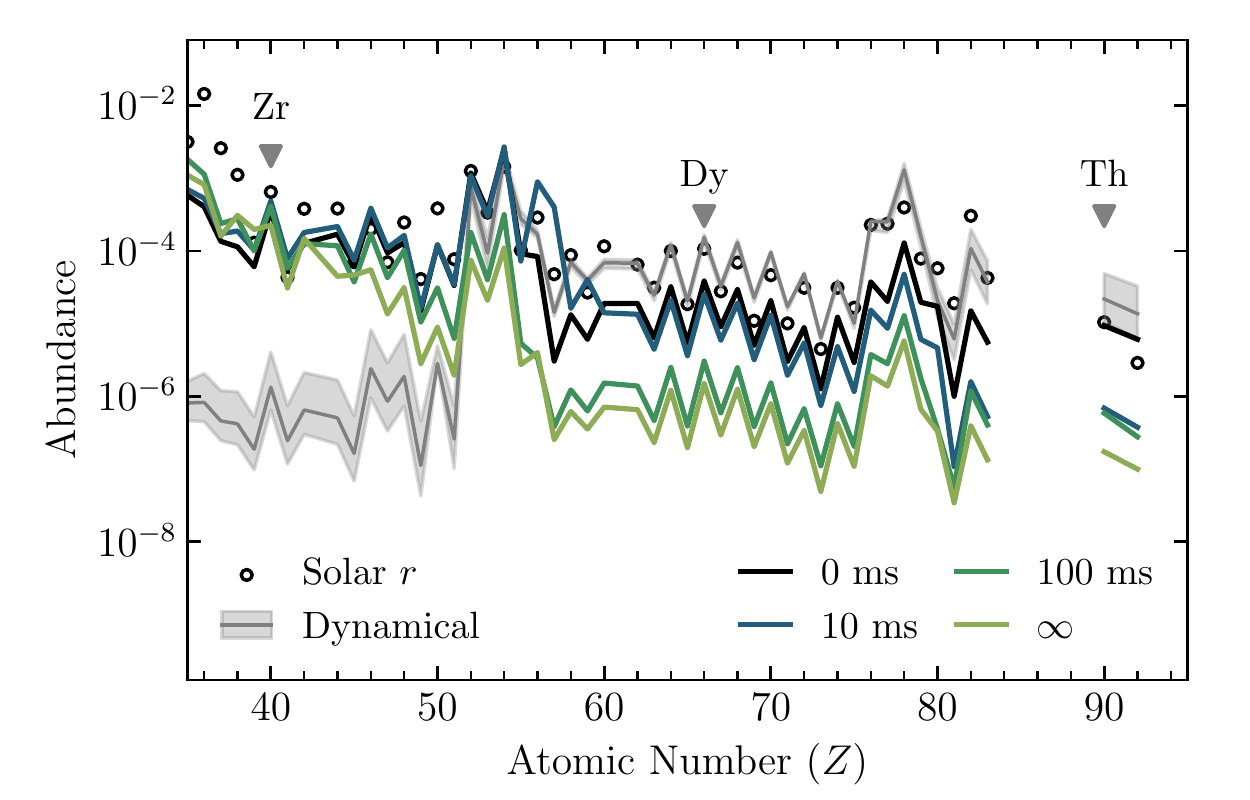}
    \includegraphics[width=\columnwidth]{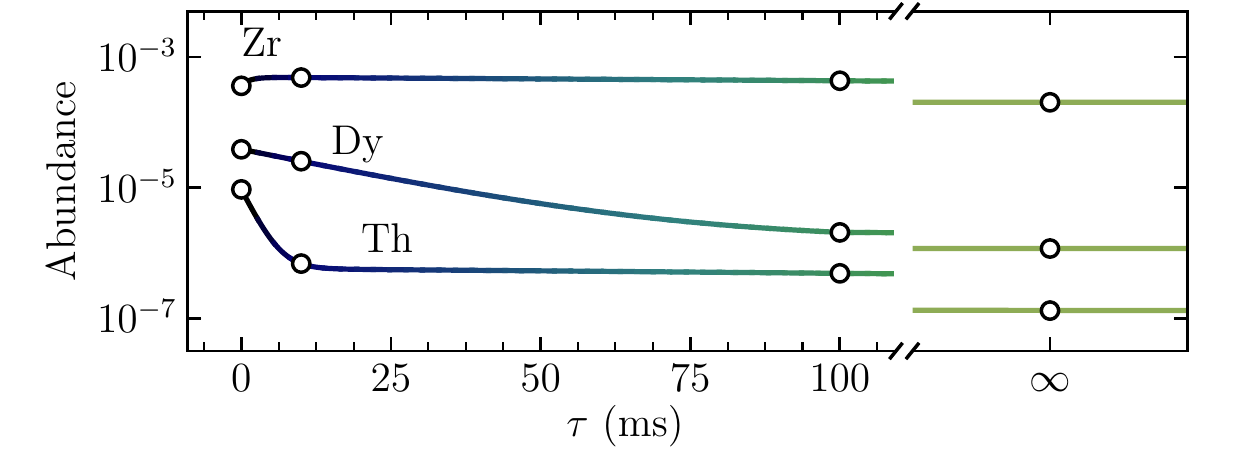}
    \caption{Top: elemental abundance patterns for the dynamical ejecta (gray) and disk outflows from HMNS merger remnants with different lifetimes (colored lines). Bottom: exponential fits to the Zr, Dy, and Th abundances versus remnant lifetime for the disk outflows.\label{fig:tau}}
	\end{figure}

For the \rp\ extent of the disk wind, we start with the same set of models used in \citet{lippuner2017}: from \citet{metzger2014} for a HMNS remnant with lifetimes 0\,ms, 10\,ms, 100\,ms, and infinitely lived.
Of roughly 10,000 tracer particles, only a fraction will be ejected in the disk outflows, depending on the lifetime of the remnant.
The tracers are each evolved with the nuclear reaction network code Portable Routines for Integrated nucleoSynthesis Modeling (PRISM; \citealt{sprouse2019}).
We start with all reactions from the JINA Reaclib database, then supply to PRISM theoretically computed \citep[from the FRDM2012 nuclear model;][]{moller2012}, experimentally evaluated, and laboratory measured data to supplement and/or overrule Reaclib input, following the same procedure as in, e.g., \citet{mumpower2018,holmbeck2019b,holmbeck2019,vassh2019}.
We employ a ``50/50" fission yield prescription as in, e.g., \citet{holmbeck2019b,vassh2019,vassh2020}. 
The details of fission fragment deposition is expected to only minimally influence the late-lanthanides (Dy--Lu) and will only affect trajectories that are neutron-rich enough for fission cycling to occur.
Astrophysically, the temperature and density evolution initially follow the simulation data.
Once the temperature of the tracer drops below 5 GK, nuclear reheating is computed by PRISM at 50\% efficiency, and the increase in entropy due to reheating is used to dynamically recalculate the temperature.
Initial compositions of the ejecta for each tracer particle are computed in nuclear statistical equilibrium (NSE) using the $Y_e$ and density from the simulation data at 10~GK.
Figure~\ref{fig:tau} shows the final, combined abundance patterns produced by a massive remnant that survives collapse for different lengths of time.

The $\log_{10}$ Zr, Dy, and Th abundances for remnants with lifetimes 0\,ms, 10\,ms, 100\,ms, and $\infty$ are well described by decaying exponential functions.
Note that decaying exponentials also fit the abundance data in \citet{lippuner2017} (see values in Table~2 therein).
Therefore, to obtain the abundances at any given lifetime, we use an exponential-based function to interpolate between the computed points.
This smooth interpolation allows us to obtain the abundances of each element at any given value of $\tau$.
These fits are shown in the bottom panel of Figure~\ref{fig:tau}.
Since the abundances change very little between 100\,ms and the infinite case, winds from remnants with lifetimes $\tau\geq300$\,ms are assumed to obey the abundances of the infinite case.

For the dynamical ejecta, we choose a trajectory from the NSM simulations of S.\ Rosswog \citep{rosswog2013,piran2013}, as in \citet{korobkin2012}.
We use PRISM in the same way as described above and allow some variations in the $Y_e$ as in \citet{holmbeck2019}.
Representative abundance patterns for very neutron-rich dynamical ejecta with $Y_e$ between 0.16 and 0.18 are shown Figure~\ref{fig:tau} compared to the wind abundance patterns calculated from the \citet{metzger2014} data.
Whereas the abundance patterns of the disk outflow depend sensitively on $M_1$ and $M_2$, we take the relative composition of the dynamical ejecta to remain extremely neutron-rich for any choice of NS masses \citep[see, however, the recent work of][]{nedora2020}.

\subsection{Stellar Sample}

\begin{deluxetable*}{rllrrrrl}
\tablecaption{List of stars with Zr, Dy, and Th abundances that we use for this work. All bracket abundances use the Solar abundances from \citet{asplund2009} for comparison.\label{tab:stars}}
\tablehead{
\colhead{ID} & 
\colhead{Star} & 
\colhead{Class\tablenotemark{a}}& 
\colhead{[Fe/H]} & 
\colhead{[Eu/Fe]} & 
\colhead{[Zr/Dy]} & 
\colhead{[Th/Dy]} & 
\colhead{Reference}
}
\startdata
  1 &  HD~110184              & \ro  &  $-$2.51  &  0.10  &  0.21  &   0.02 & \citet{honda2004} \\ 
  2 &  BD\,+10$^\circ$2495    & \ro  &  $-$2.31  &  0.12  & $-$0.16  &   0.17 & \citet{roederer2010} \\ 
  3 &  BD\,$-$15$^\circ$5781  & \ro  &  $-$2.92  &  0.14  & $-$0.20  &  $-$0.06 & \citet{roederer2014} \\ 
  4 &  J08580584$-$0809174    & \ro, Ac$-$&   $-$3.16  &  0.25  & $-$0.20  &  $-$0.16 & \citet{cain2018} \\ 
  5 &  HD~186478              & \ro  &  $-$2.75  &  0.25  & $-$0.15  &   0.08 & \citet{roederer2014} \\ 
  6 &  HD~108577              & \ri  &  $-$2.36  &  0.38  & $-$0.31  &   0.08 & \citet{johnson2001,johnson2002} \\ 
  7 &  BD+08$^\circ$2856      & \ri  &  $-$2.10  &  0.44  & $-$0.45  &  $-$0.12 & \citet{johnson2001,johnson2002} \\ 
  8 &  HD~6268                & \ri  &  $-$2.82  &  0.45  & $-$0.42  &  $-$0.04 & \citet{roederer2014} \\ 
  9 &  HD~108317              & \ri, Ac$-$  &  $-$2.53  &  0.64  &  $-$0.28  &  $-$0.16  &  \citet{roederer2012}\\
 10 &  CS~30315-029  &  \ri, Ac$+$  &  $-$3.43  &  0.69  & $-$0.63  &   0.25 & \citet{siqueira2014} \\ 
 11 &  HD~221170  &  \ri  &  $-$2.16  &  0.80  & $-$0.48  &  $-$0.06 & \citet{ivans2006} \\ 
 12 &  HD~115444  &  \ri, Ac$-$  &  $-$2.96  &  0.83  & $-$0.49  &  $-$0.14 & \citet{westin2000} \\ 
 13 &  BD\,+17$^\circ$3248  &  \ri  &  $-$2.06  &  0.89  & $-$0.69  &  $-$0.07 & \citet{cowan2002} \\ 
 14 &  CS~30306-132  &  \ri, Ac$+$  &  $-$2.41  &  0.89  & $-$0.40  &   0.39 & \citet{honda2004} \\ 
 15 &  CS~22953-003  &  \ri  &  $-$3.13  &  0.94  & $-$0.79  &   0.03 & \citet{roederer2014} \\ 
 16 &  J20050670$-$3057445  &  \ri, Ac$-$  &  $-$3.03  &  0.96  & $-$0.76  &  $-$0.18 & \citet{cain2018} \\ 
 17 &  HE~2327-5642  &  \ri  &  $-$2.78  &  0.99  & $-$0.84  &   0.01 & \citet{mashonkina2010} \\ 
 18 &  CS~29491-069  &  \rii, Ac$-$  &  $-$2.60  &  1.07  & $-$0.74  &  $-$0.13 & \citet{hayek2009} \\ 
 19 &  HE~2252$-$4225  &  \rii  &  $-$2.94  &  1.14  & $-$0.83  &   0.03 & \citet{mashonkina2014} \\ 
 20 &  J15383085$-$1804242  &  \rii, Ac$-$  &  $-$2.09  &  1.27  & $-$0.90  &  $-$0.20 & \citet{sakari2018b} \\ 
 21 &  COS82                &  \rii, Ac$-$  &  $-$1.47  &  1.29  &  $-$0.79  &  $-$0.30 &  \citet{aoki2007} \\
 22 &  J09544277+5246414    &  \rii, Ac$+$  &  $-$2.99  &  1.30  & $-$0.95  &   0.43 & \citet{holmbeck2018} \\ 
 23 &  HE~1219$-$0312       &  \rii  &  $-$2.97  &  1.41  & $-$0.86  &   0.13 & \citet{hayek2009} \\ 
 24 &  J14325334$-$4125494  &  \rii  &  $-$2.97  &  1.46  & $-$0.96  &   0.00 & \citet{cain2018} \\ 
 25 &  CS~31082-001  &  \rii, Ac$+$  &  $-$2.90  &  1.62  &  $-$0.84  &  0.31  &  \citet{hill2002} \\
 26 &  CS~22892-052  &  \rii, Ac$-$  &  $-$3.10  &  1.65  & $-$1.02  &  $-$0.26 & \citet{sneden2003} \\ 
 27 &  J20384318$-$0023327  &  \rii  &  $-$2.91  &  1.66  & $-$0.78  &   0.17 & \citet{placco2017} \\ 
 28 &  CS~29497-004  &  \rii, Ac$-$  &  $-$2.68  &  1.73  & $-$0.88  &  $-$0.20 & \citet{christlieb2004} \\ 
 29 &  J033523$-$540407  &  \rii, Ac$-$  &  $-$3.05  &  1.78  & $-$0.99  &  $-$0.47 &  \citet{ji2018} \\
\enddata
\tablenotetext{a}{Both the \rp-enhancement and the actinide-boost classes are noted, with ``Ac$-$" indicating an actinide-deficient signature and ``Ac$+$" an actinide-boost, following the definitions presented in \citet{holmbeck2019b}. No label indicates actinide-normal stars.}
\end{deluxetable*}

The aim of this study is to explore how observed stellar \rp\ signatures can be explained with theoretical descriptions of NSM outflows.
We choose a subset of metal-poor stars with \rp\ abundances and hypothesize that most of their actinides were produced in NSM dynamical ejecta, and, conversely, most of their \limr\ elements were produced in NSM disk outflows.

We choose the Th/Dy abundance ratio to represent the relative production of actinide-to-main \rp\ material and the Zr/Dy ratio to represent limited-to-main \rp\ production.
These three elements and their ratios capture the relevant regions of the abundance pattern to provide a sense of \rp\ extent in an astrophysical environment.
We select all metal-poor stars in JINAbase that have reported measurements of Zr, Th, and Dy (29 in total), listed in Table~\ref{tab:stars}.
These stars vary from non-enhanced (``\ro": $[{\rm Fe/H}]\leq 0.3$), to slightly enhanced (\ri), to very enhanced (\rii), based on the definitions in \citet{frebel2018}.
Although the \ro\ stars with low \rp\ levels are not technically defined as ``\emph{enhanced}," their low \rp\ levels might simply arise from a larger Fe-polluted gas mass.

Next, we describe how we use the observational Zr, Th, and Dy abundances of these 29 metal-poor stars as input to the MCMC method to derive progenitor NS binary masses.
For this study, we work under the assumption that the majority of \rp\ material---especially second-peak elements and beyond---originated from one \rp\ event whose outflows were efficiently mixed into the ISM, and therefore that each star can be traced to individual NSMs.

\subsection{ADMC: The MCMC Method}

\begin{figure}[ht]
	\includegraphics[width=\columnwidth]{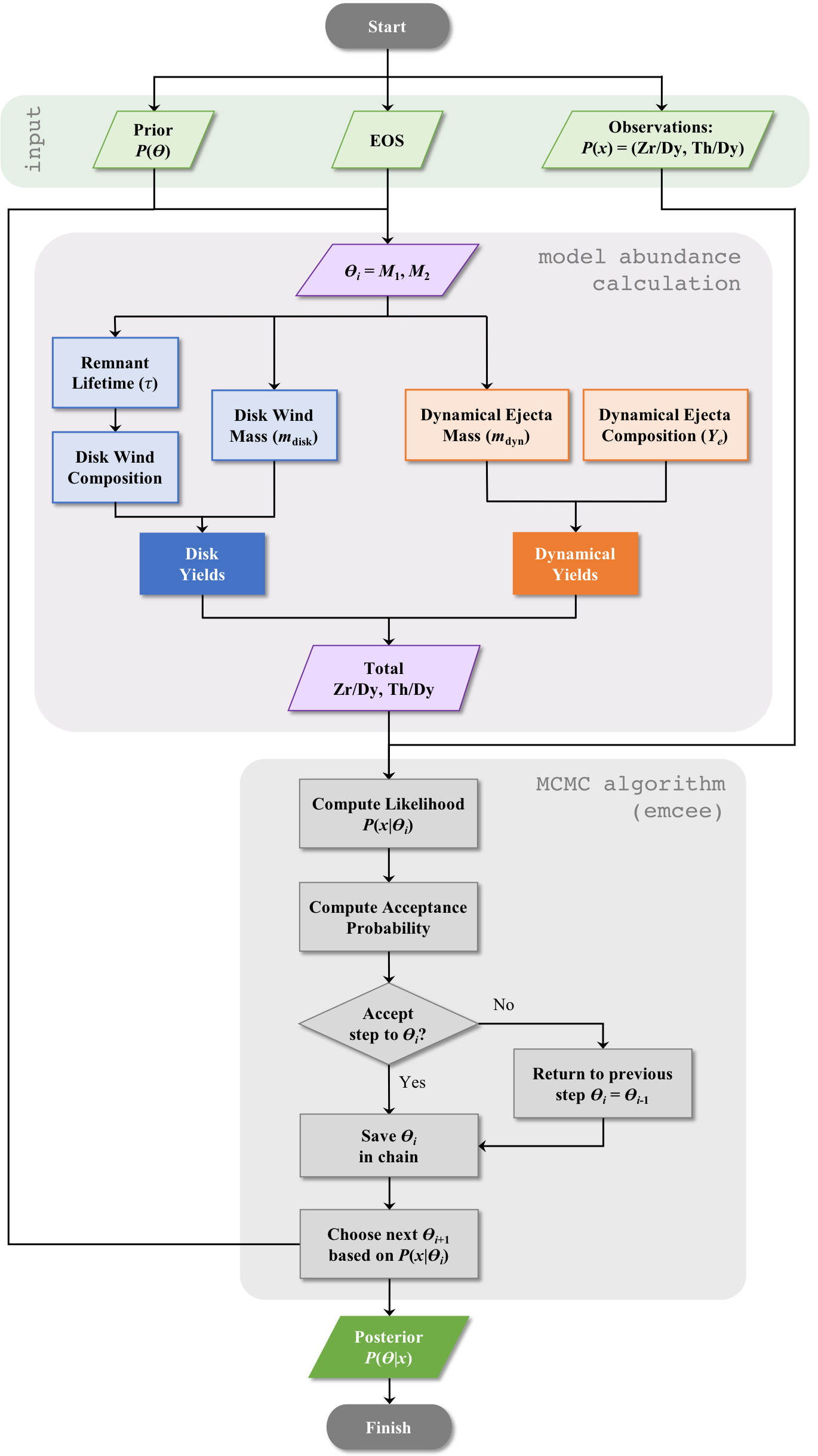}
	\caption{Schematic flow of how all the ADMC model inputs map to the output total abundances and produce a posterior distribution of likely $M_1$ and $M_2$ solutions.
	\label{fig:flowchart}}
\end{figure}

To reconstruct the merger properties of the systems that could have produced the \rp-enhanced stars, we use the \texttt{emcee} Python package \citep{foreman2013}, a Markov chain Monte Carlo (MCMC) algorithm, to explore $M_1$ and $M_2$ combinations of the progenitor NS binary.
The way in which we apply this method takes observational abundance ratios as input and generates distributions of predicted $M_1$ and $M_2$ as output.
Figure~\ref{fig:flowchart} shows a simple visualization of how the set of input propagates through to a likelihood calculation and posterior distribution of $M_1$ and $M_2$ solutions.

First, for each EOS, the MCMC randomly samples $M_1$ and $M_2$ from a supplied prior distribution.
Then, with this random choice of $M_1$ and $M_2$, we apply Equations~\ref{eqn:disk} and \ref{eqn:dynk} to determine how much mass is lost due to the wind outflow and dynamical ejecta mechanisms, yielding $m_{\rm wind}$ and $m_{\rm dyn}$, respectively.

These $M_1$ and $M_2$ values simultaneously provide a unique estimate of the lifetime of the NSM remnant before it collapses into a black hole from Equation~\ref{eqn:tau}, which in turn sets a characteristic abundance pattern of the wind ejecta (see Figure~\ref{fig:tau}).
We take the model Zr, Dy, and Th abundances obtained from the fitted wind outflow patterns for any remnant lifetime, then scale the abundances by $m_{\rm wind}$ found previously.
Lastly, these abundances are multiplied by 25\% to account for only a fraction of the total disk mass being ejected.
After these steps, we obtain the predicted total Zr, Dy, and Th ejected by the disk outflows using $M_1$, $M_2$, and EOS as input.

The computed abundance pattern of the dynamical ejecta is similarly scaled by $m_{\rm dyn}$.
The relative abundances produced in the dynamical ejecta is mostly insensitive to the NSM properties; still, the initial $Y_e$ of dynamical ejecta is unknown.
We let the MCMC algorithm also sample $Y_e$ and find that the range 0.16--0.18 can account for almost all input stellar abundances.
Therefore, for a particular choice of $M_1$ and $M_2$, a total ejected Zr, Dy, and Th abundance from two ejecta mechanisms within the merger are obtained by first finding $m_{\rm wind}$ and $m_{\rm dyn}$, then multiplying representative abundance patterns of the wind and dynamical ejecta by the corresponding ejecta masses.
In this way, we obtain the total mass yields of \rp\ ejecta from the NSM event.

The algorithm computes output abundances from $M_1$ and $M_2$ for each EOS as it explores the $M_1$-$M_2$ parameter space, preferring solutions whose computed abundance ratios are within certain tolerances on the input observed Zr/Dy and Th/Dy abundances.
Our MCMC algorithm uses 50 ``walkers" distributed across the $M_1$-$M_2$ parameter space within theoretical limits of NS masses (i.e., $M_{1,2}\leq M_{\rm TOV}$ and $M_1 \geq M_2$).
The walkers are distributed according to a prior that is fitted to the binary NS distribution: essentially a Gaussian with a peak near 1.35\,M$_\odot$ and a standard deviation of $\sim$0.1\,M$_\odot$, obtained from fitting to data in \citet{tauris2017}.
\citet{ozel2012} also find a similar fit to these data.
Implementing a prior that corresponds to current estimates of DNS mass distributions is a source of uncertainty in our method.
At least some of the stellar abundances we are using are expected to have been produced early in the evolution of the universe, and it is unclear if or how the distribution of DNS masses has evolved since then.
From initial investigations, we find that the MCMC results are generally insensitive to the assumption that the DNS distribution has not changed over time.
In Section~\ref{sec:noprior} we will the discuss in detail how changing the prior has the potential to alter our results.

Using Equations~\ref{eqn:disk}--\ref{eqn:tau}, the ratios of the combined Zr/Dy and Th/Dy model abundances are compared to the observationally derived abundances to determine the likelihood that a particular $M_1$-$M_2$ combination reproduces the input abundance ratios obtained from observations.
The tolerances supplied to the likelihood function are such that the fitted Zr/Dy be at most 0.1~dex greater than the input Zr/Dy abundance, but allow the Zr/Dy to be significantly lower.
We choose a relaxed constraint on Zr/Dy since some amount of Zr could come from SNe sites that also created the Fe-peak elements present in the star.
However, we require that the Th/Dy is at maximum $\pm$0.1~dex from the input value, since we presume that the majority of main \rp-element abundances in each star were created by a merger event.
As a result, the MCMC algorithm---which we now name ADMC (Actinide-Dilution with Monte Carlo)---finds a distribution of probable combinations of $M_1$ and $M_2$ of the original NS binary that merged to produce nearly all the Zr, Dy, and Th abundances present in each stellar atmosphere.

	\begin{figure}[t!]
	\centering
    \includegraphics[width=\columnwidth]{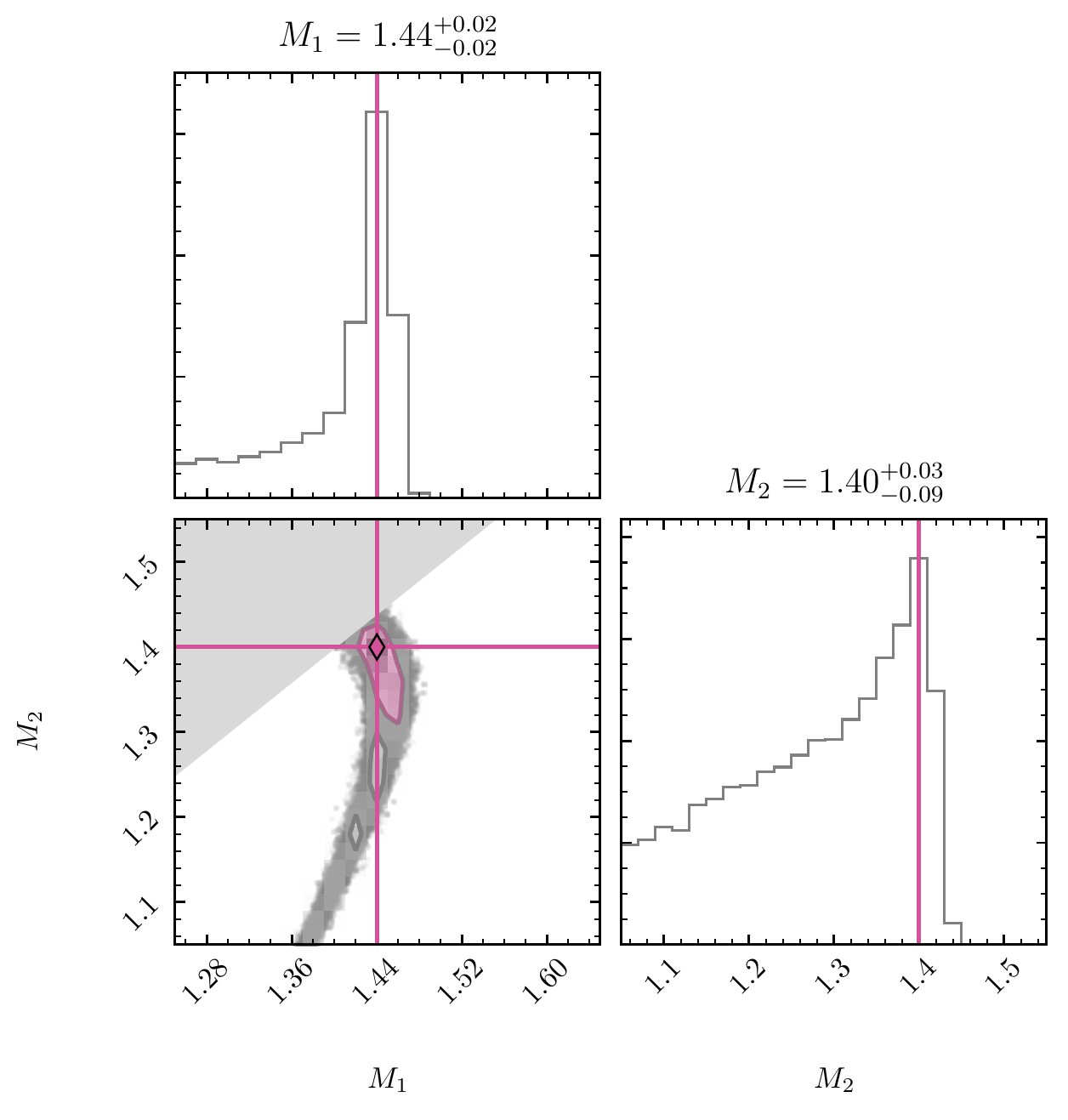}
    \caption[Corner plot showing a sample ADMC result for the most likely combination of $M_1$ and $M_2$]{Corner plot showing the most likely combination of $M_1$ and $M_2$ for CS30306-132 (Star ID 14) using the DD2 EOS. The contour of 1-$\sigma$ solutions is shaded in pink.\label{fig:corner}}
	\end{figure}

Figure~\ref{fig:corner} shows an example of the mass results of ADMC for CS30306-132 (Star ID 13) using the DD2 EOS.
Due to the small parameter space, we found that 2000 steps with a burn-in stage of 500 is more than sufficient for the walkers to thoroughly explore the parameter space.
The gray shaded region in the upper-left corner of the $M_1$-$M_2$ plot shows one of the limits supplied as a prior: $M_2$ will always be less than or equal to $M_1$.
Without this constraint, ADMC would simply find superfluous solutions reflected about the $M_1=M_2$ axis, which are equivalent to $M_{1\leftrightarrow 2}$.
The long tail visible in the $M_1$-$M_2$ plot offers a view into the complexity of the parameter space and an indication that $M_1$ and $M_2$ are not independent of each other.
Under the constraints we use with ADMC, only delicate combinations of NS masses can reproduce the input abundances within the supplied tolerances.
The pink colored region in Figure~\ref{fig:corner} denotes 1-$\sigma$ edges of the solutions in the two-dimensional parameter space.
Reported 1-$\sigma$ uncertainties on $M_1$ and $M_2$ are taken from the limits of these asymmetric contour edges.

\section{Results}
\label{sec:results}

We now discuss our results by comparing our theoretically derived NS masses with observations of present-day NSs in the Galaxy.
The ADMC results provide for each input star (i.e., pair of observationally derived Zr/Dy and Th/Dy abundances) a distribution of the primary ($M_1$) and secondary ($M_2$) NS masses.
These $M_1$ and $M_2$ combinations are NS masses whose computed merger outflow compositions reproduce the observed (input) stellar abundances.
First, we will discuss the individual mass results for the 29 stars in Table~\ref{tab:stars}.
Then, we will attempt to broaden our result to a larger population of stars in order to compare to Galactic distributions of NS systems.

	\begin{figure*}[pt]
	\centering
    \includegraphics[width=0.325\textwidth]{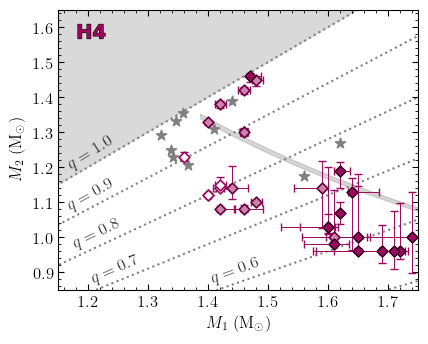}
    \includegraphics[width=0.325\textwidth]{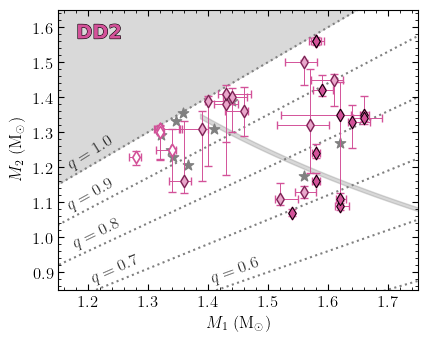}
    \includegraphics[width=0.325\textwidth]{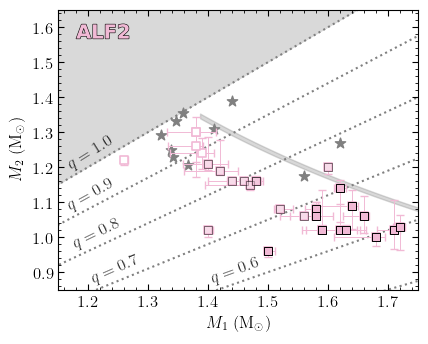}\\
    \includegraphics[width=0.325\textwidth]{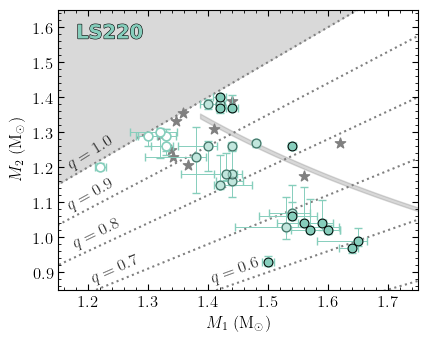}
    \includegraphics[width=0.325\textwidth]{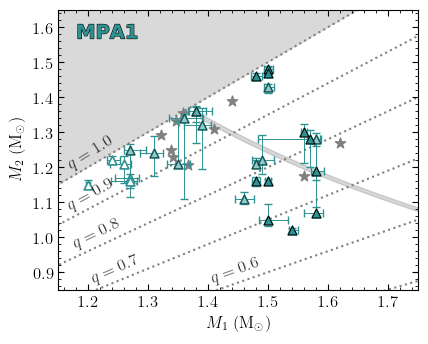}
    \includegraphics[width=0.325\textwidth]{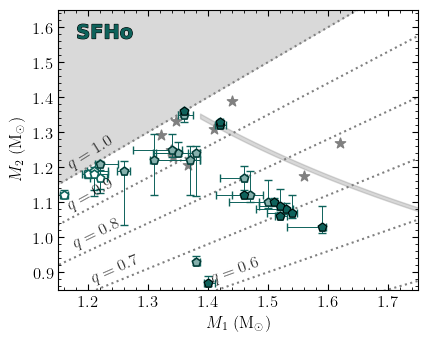}
    \caption{Mass results for each EOS choice compared to known DNS systems (stars). Lines of constant $q$ are shown for comparison. Dark-filled symbols with black outlines denote \rii\ stars, light-filled symbols denote the \ri\ stars, and white-filled symbols with colored outlines indicate \ro\ stars. The shaded band in the upper-right shows the range of NS masses satisfying the chirp mass of GW170817: $\mathcal M=1.188_{-0.002}^{+0.004}$\,M$_{\odot}$.\label{fig:m1m2}}
	\end{figure*}


Figure~\ref{fig:m1m2} shows the most likely $M_1$-$M_2$ pairs that would have merged to account for the observed \rp\ abundances of each star in Table~\ref{tab:stars} for each EOS. 
These ADMC-predicted binary progenitors are colored by the distinct \rp\ enhancement signature that their outflows reproduce: \rii\ (dark-filled symbols), \ri\ (lightly filled symbols), or non-enhanced (white-filled). We also compare our results with precisely measured $M_1$ and $M_2$ pairs obtained from present-day DNS systems in the Galaxy (gray stars).
Lines of constant $q$ are shown for guidance (dotted), and the diagonal thin gray band displays the range of $M_1$-$M_2$ solutions that satisfy the chirp mass of GW170817: $\mathcal M=1.188_{-0.002}^{+0.004}$\,M$_{\odot}$. Although not pictured in the plot, lines of constant total mass ($M_{\rm tot}$) would run diagonally from top-left to bottom-right, roughly parallel to the GW17817 chirp mass band.
We also list the individual mass results by EOS in the tables in Supplemental Materials. Since our MCMC tool produces a posterior {\it distribution} of solutions, the asymmetric error bars on each $M_1$-$M_2$ pair show the limits of the two-dimensional 1-$\sigma$ contours from the ADMC output (see Figure~\ref{fig:corner}). These uncertainties correspond directly to the provided tolerance on how far the computed abundances are allowed to deviate from the input abundances and have the MCMC still consider the $M_1$-$M_2$ pair producing those abundances a successful match.

Before we discuss our results further, we note that the ADMC output yields and their interpretation rely on several underlying assumptions:
\begin{enumerate}
	\item The majority of \rp\ material in the Galaxy was made by NSMs. 
    \item The \rp\ abundances in each star in our study are the result of only one major NSM progenitor, and the elemental production by each of these events extends over the entire \rp\ pattern from the \limr\ elements (Zr) all the way to the actinides (Th).
	\item The dynamical ejecta are extremely neutron-rich and contribute nearly no Zr abundance to the combined abundances of the total ejecta. Similarly, the composition of the disk wind is determined by simulations with no additional neutrino or magneto-hydrodynamic effects altering their composition.
	\item The mass distribution of DNS systems is roughly constant across cosmic time, so that the present-day distribution of DNSs in the Galaxy may be compared with earlier distributions.
\end{enumerate}

\subsection{Trends with Enhancement Level}

Now we will discuss in detail trends that appear in our results and their possible origins within our MCMC framework.
The first noticeable effect in the ADMC results is the trend with stellar \rp\ enhancement level.
As can be seen in Figure~\ref{fig:m1m2}, NSM progenitors of the \rii\ stars appear to be predominantly asymmetric NS binary systems, when using all the above assumptions.
For stars with \ri\ signatures, ADMC predicts slightly more symmetric progenitor binary systems, and the abundances of \ro\ stars are well reproduced by nearly symmetric---and quite low-mass---NS binaries.

These results can be explained by the observational trends of first \rp\ peak elements to main \rp\ elements in metal-poor stars.
The \rii\ stars have a tendency to be more metal-poor and display lower Zr/Dy ratios than the typically less metal-poor \ri\ stars that tend to exhibit higher Zr/Dy values.
While the exact origin of this observed difference is unknown, there is clearly more nucleosynthetic information hidden in this ratio that could unlock further clues about other progenitors in combination with erstwhile \rp\ sites.
For this analysis, we assume that all Zr/Dy originated from one \rp\ site.
Because of this requirement, our MCMC method assigns symmetric systems to high Zr/Dy stellar abundances (\ro\ and \ri\ stars), and asymmetric solutions to low Zr/Dy ratios (primarily \rii\ stars).
To match a low Zr/Dy abundance ratio, the total Zr produced by the disk and dynamical ejecta must be relatively low compared to the total Dy.
However, the Zr yield is almost exclusively produced by the disk outflows, which also contribute some non-negligible amount of Dy.
Therefore, one way to obtain a low Zr/Dy ratio is to increase the Dy yield by the dynamical ejecta, which contributes nearly no additional Zr.
This increase in dynamical ejecta can be accomplished by an increase to the NS binary mass asymmetry, hence explaining why the MCMC predictions for \rii\ stars are primarily at low $q$ values.

\subsection{EOS Effects}

Now we discuss the results as obtained for the different EOSs.
The dependence on the EOS manifests mainly through $M_{\rm TOV}$ and how the maximum NS mass plays into the total disk ejecta.
For EOSs with a lower $M_{\rm TOV}$ (i.e., H4, ALF2, LS220, and SFHo), the results in Figure~\ref{fig:m1m2} tend to gather along values of constant total binary mass: diagonals running roughly parallel to the GW170817 chirp mass.
Recall that the total binary mass, $M_{\rm tot}$, directly influences the total mass of the disk outflows as well as the collapse time of the remnant into a black hole.
Both the disk mass and the collapse lifetime then determine the majority of Zr production primarily in disk wind in the NSM ejecta.
Therefore, the grouping of our results along certain values of $M_{\rm tot}$ reflects the need to match the observed Zr/Dy ratios.

As discussed above, many of the \rii\ and highly enhanced \ri\ stars have similar, low Zr/Dy abundance ratios.
Since the Zr/Dy is primarily set by the disk outflows---in turn governed by the total mass---the solutions for these stars therefore favor similar values of $M_{\rm tot}$.
The relative Th/Dy abundances of the \ri\ and \rii\ stars differ much more broadly.
Therefore the optimal $M_1$-$M_2$ solution matching the input abundances is found at different values of $q$ along comparable $M_{\rm tot}$ values, explaining why many of the results appear strung along a diagonal line in Figure~\ref{fig:m1m2}.

For many of our predicted NSMs, the amount of mass ejected by the disk wind outflows is roughly constant for most EOSs, while the amount of dynamical ejecta varies widely.
This wide variation can be seen in the tables in Supplemental Materials.
For many cases, the variation can be accounted for by a low-mass primary as described above, explaining why so many solutions in Figure~\ref{fig:m1m2} have $M_1<1.4$\,M$_\odot$.
Hence, the observed Th/Dy abundance ratios act as a sensitive probe of \rp\ conditions, such as those in merger environments, and could potentially be a key to using observed actinide abundances as EOS constraints. Nevertheless, the large range of $m_{\rm dyn}$ also emphasizes the importance of measuring actinide abundances in metal-poor \rp\ stars.

EOSs with a larger $M_{\rm TOV}$, i.e., MPA1 and DD2, produce merger remnants that survive collapse the longest.
The effect of $M_{\rm TOV}$ on remnant collapse lifetime is shown explicitly in Equation~\ref{eqn:tau}; for systems with the same NS mass, assuming an EOS with a larger $M_{\rm TOV}$ will produce a merger that will collapse sooner.
Therefore, to achieve the same result as an EOS with a smaller $M_{\rm TOV}$, the remnant must survive for \emph{longer} before it collapses to compensate for the same level of total \limr\ (Zr) production.
However, a remnant that survives collapse for longer also produces an increased limited (or weak) \rp\ component in its wind outflows, yielding a high Zr abundance, but very little Dy.
The lack of Dy in the disk wind then necessitates that the majority of main \rp\ material (Dy) must come from the dynamical ejecta.
Consequently, many of the results arising from assuming an EOS with a large $M_{\rm TOV}$ are grouped within a narrow range of $q$ (0.9--1.0), which primarily determines the dynamical ejecta yield.
Once grouped in a narrow range of $q$ values that essentially determines the Th/Dy abundance ratio of the ejecta, the $M_1$-$M_2$ solutions are spread across a wide range of $M_{\rm tot}$ in order to then match the observed Zr/Dy abundances. 
Figure~\ref{fig:m1m2} shows the results for EOS DD2 and MPA1. They do not tend to follow lines of constant $M_{\rm tot}$ (diagonals roughly parallel to $\mathcal M$) as the results based on other EOSs with smaller $M_{\rm TOV}$s, but rather favor a narrow ranges of $q$.
We thus conclude that the EOS-dependence is more sensitive to $M_{\rm TOV}$ than the detailed mass-radius relationship (i.e., compactness) set the EOS. Accordingly, this method has potential to be used to rule out EOSs based on their $M_{\rm TOV}$ rather than, e.g., their predicted radius of a 1.4-M$_{\odot}$ NS.


\subsection{Comparison to NS Observations}

\begin{figure*}[ht]
	\centering
    \includegraphics[width=0.325\textwidth]{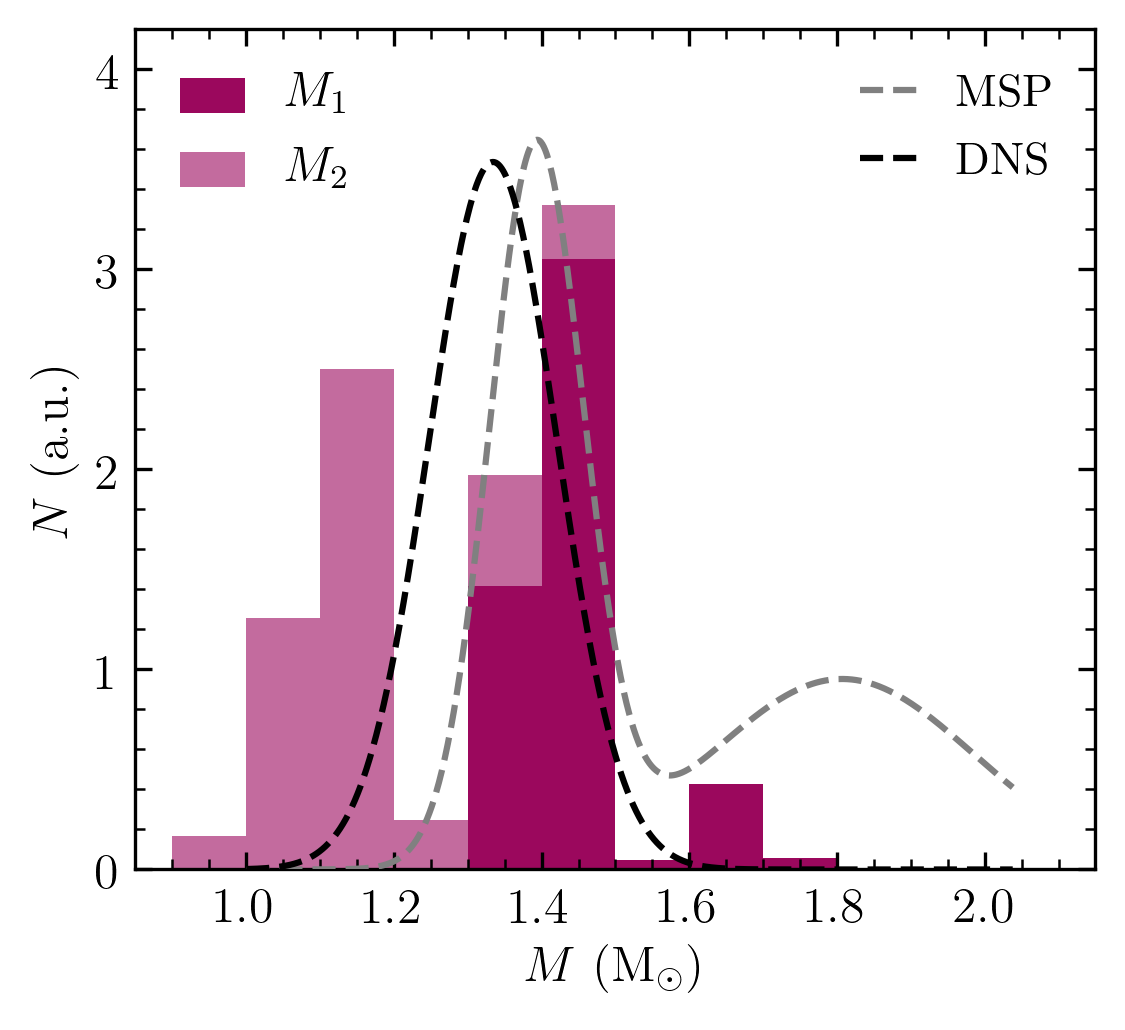}
    \includegraphics[width=0.325\textwidth]{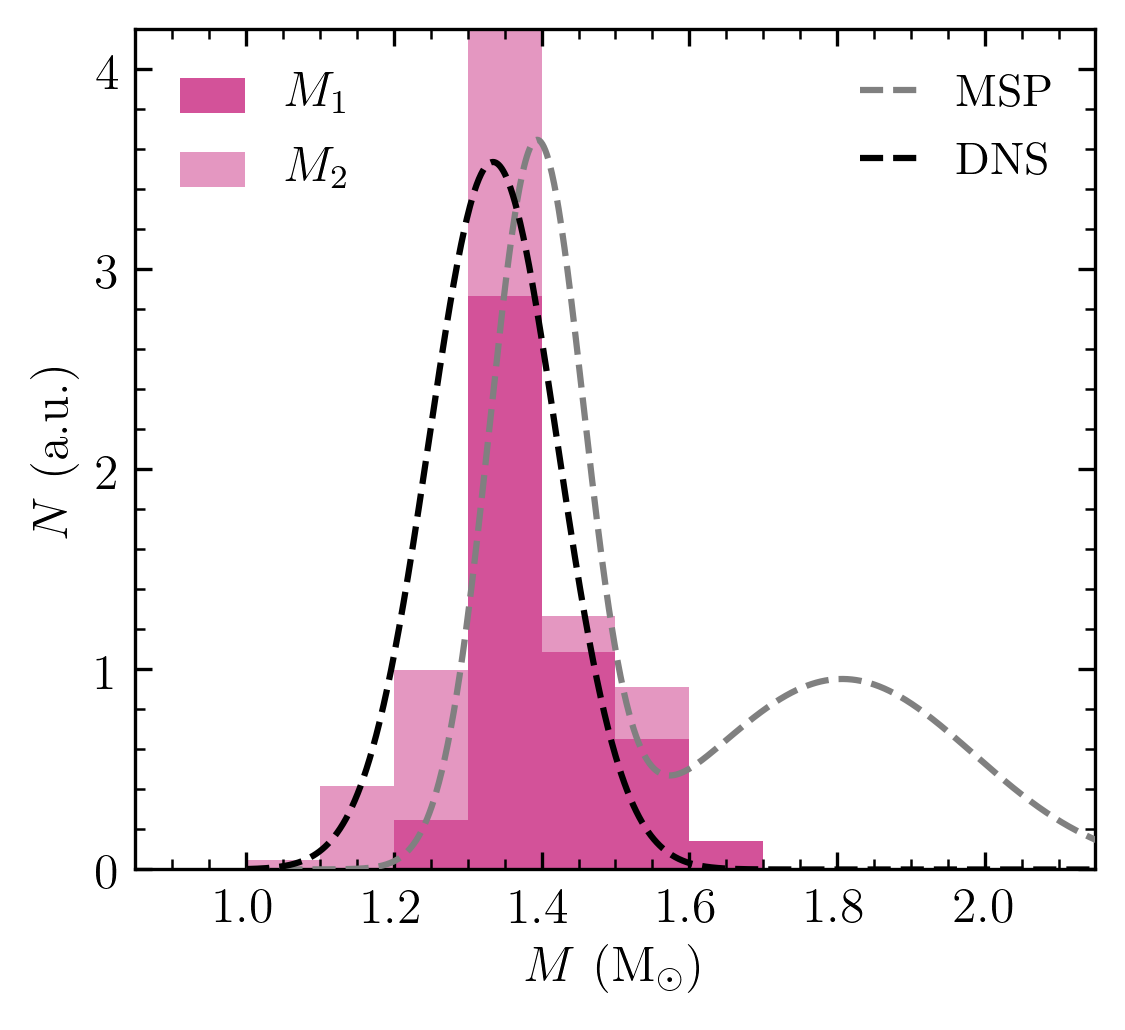}
    \includegraphics[width=0.325\textwidth]{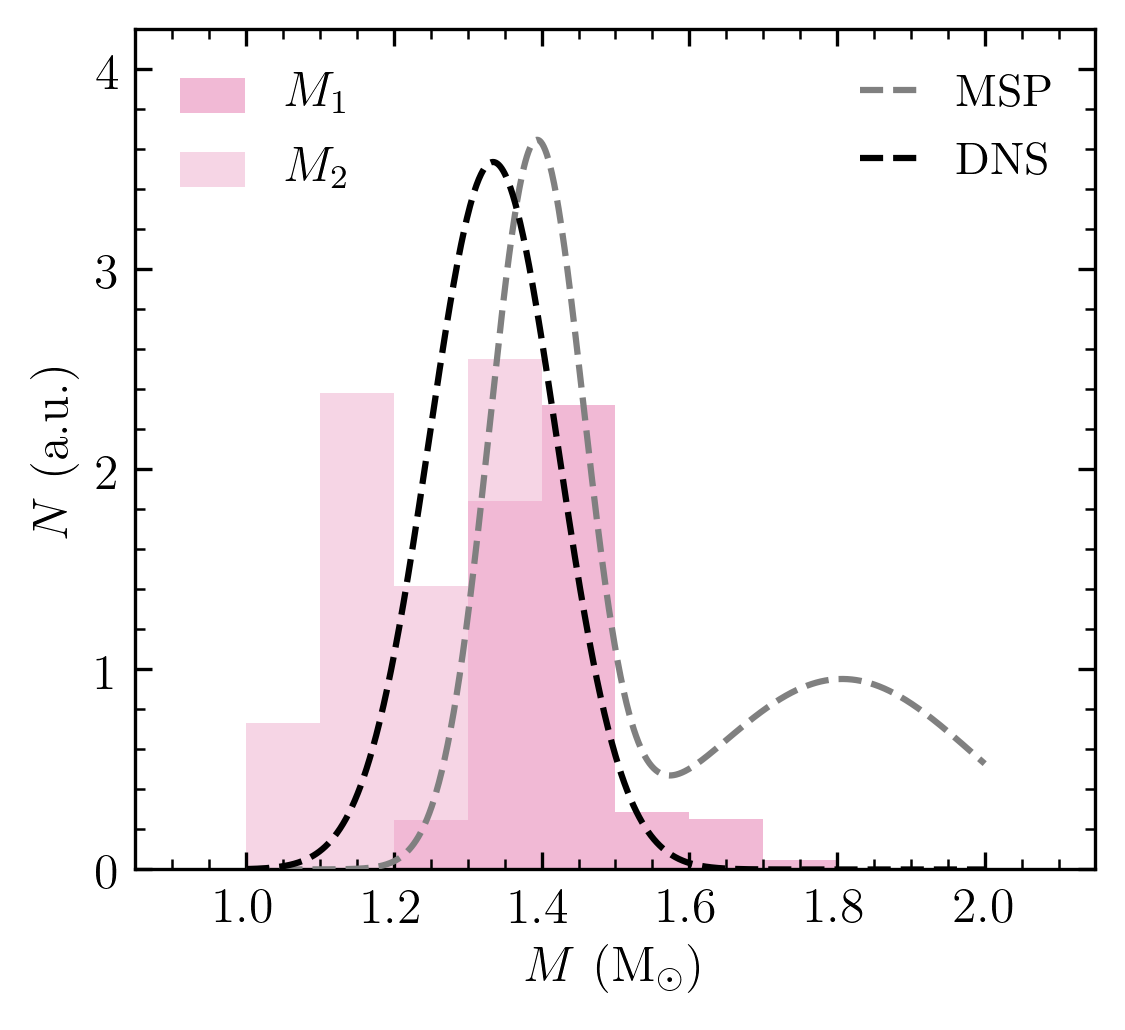}\\
    \includegraphics[width=0.325\textwidth]{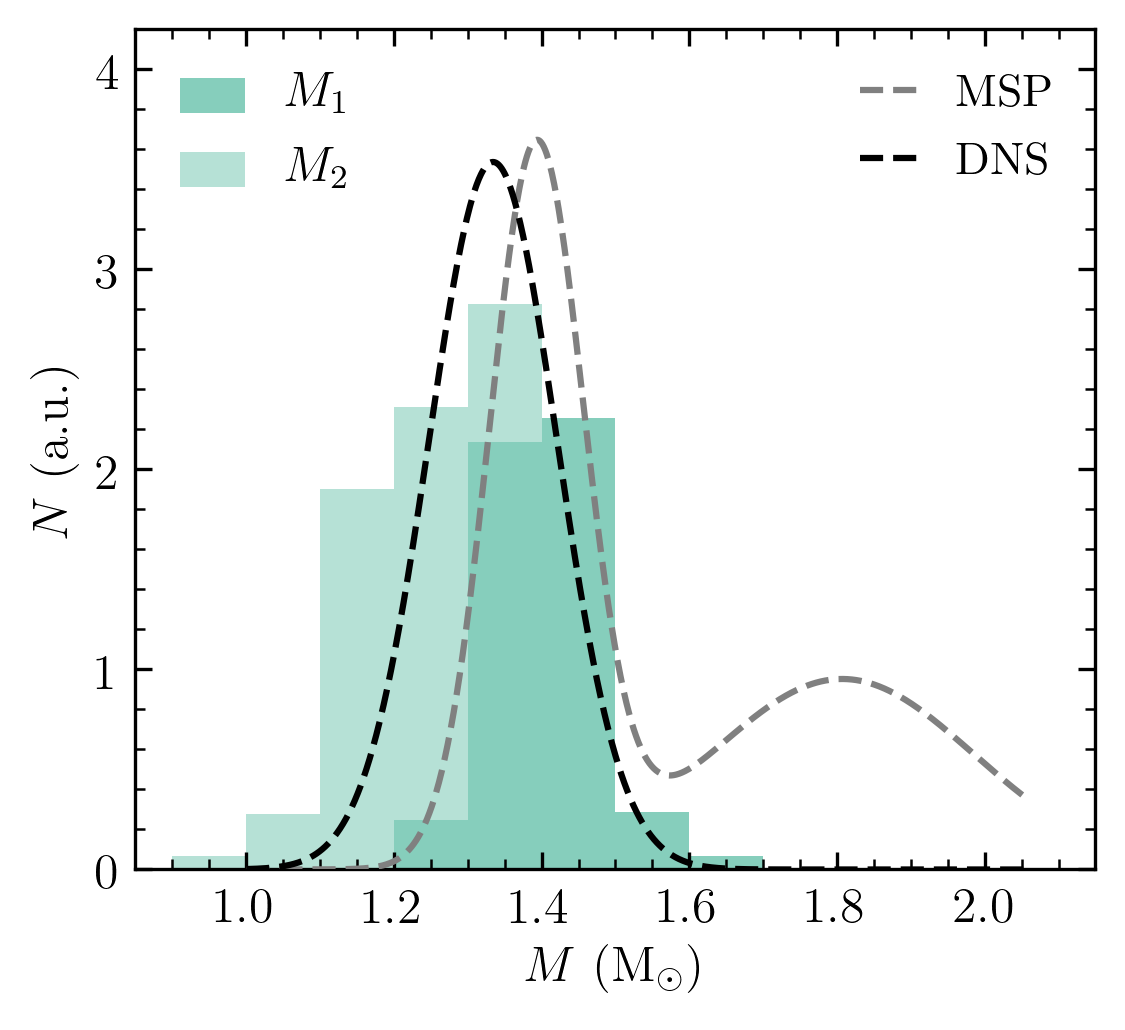}
    \includegraphics[width=0.325\textwidth]{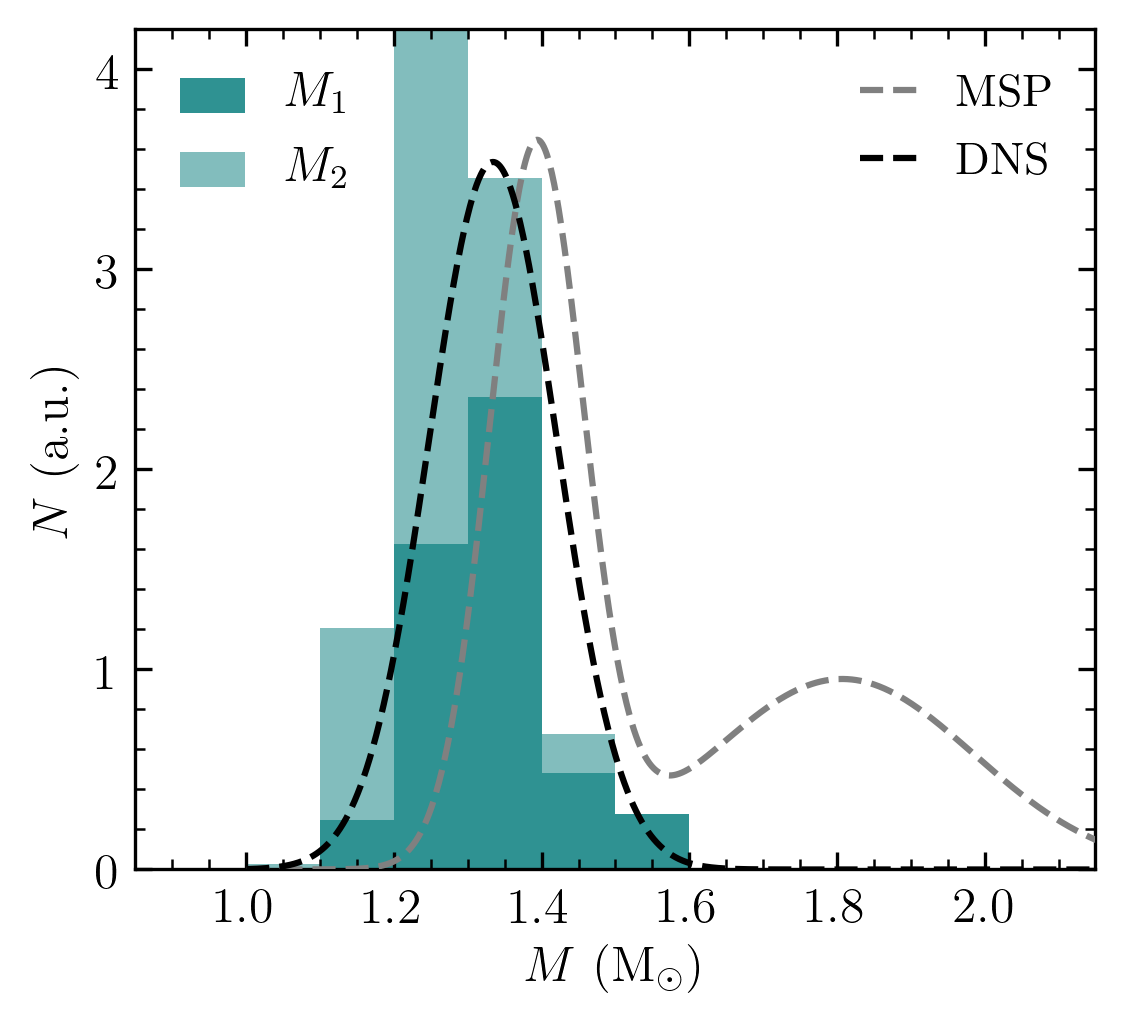}
    \includegraphics[width=0.325\textwidth]{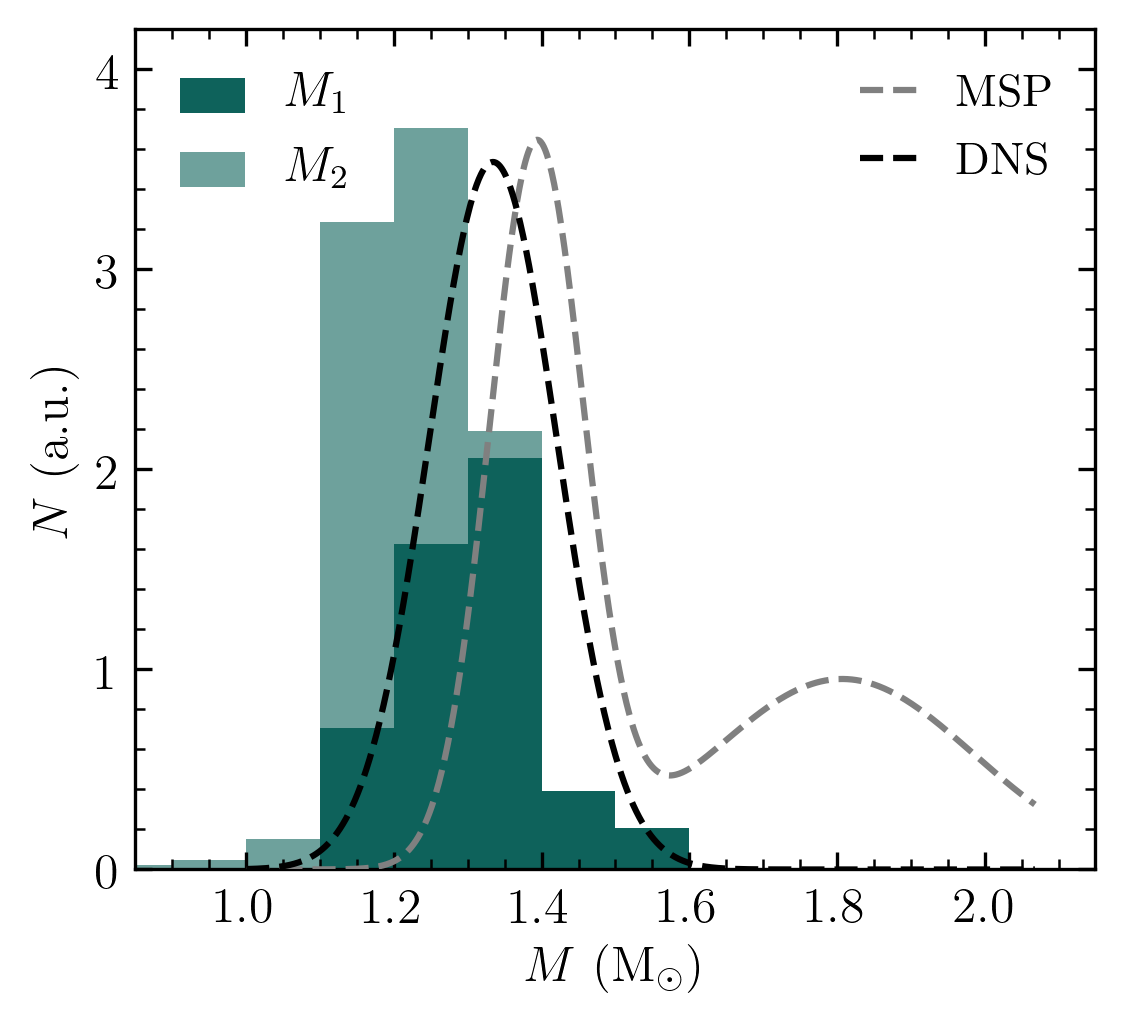}
    \caption{Scaled mass distributions for each EOS choice, weighted by the [Eu/Fe] distributions of RPA measurements. Each panel indicates separate EOSs, using the same order and colors as in Figure~\ref{fig:m1m2}. The contribution to the total mass distribution by the primary ($M_1$) and secondary ($M_2$) NSs are indicated by color shading. The gray and black dashed curves show the fitted mass distributions for NSs in DNS and MSP systems (see text for details).\label{fig:nstar}}
	\end{figure*}

The ADMC results can also be compared to observational results of present-day, individually measured NS masses.
Our 29 ADMC results must first be extended to a larger set of metal-poor stars in order to directly address the first assumption that the \emph{majority} of \rp\ material in the Galaxy was made by NSMs. 
One assumption implicitly folded in when obtaining the mass distributions (i.e., Figure~\ref{fig:nstar}) is that \rp-enhanced Galactic halo metal-poor stars are accurately represented by our 29 stars.
Clearly, given the specific selection of this sample by requiring \rp\ enhanced stars with available measurements of Zr, Dy, and Th is highly biased.
Hence, our 29 stars unfortunately cannot not well represent the body of \rp-enhanced metal-poor halo stars.

While we cannot quantify the selection function for our sample, we can attempt to account for any introduced biases (i.e., the lack of available Zr, Dy, and Th measurements) in more metal-poor stars. We do so by comparing our sample and its distribution in [Eu/Fe] to a large sample of \rp-enhanced stars that was collected in a fairly unbiased fashion. This allows us to essentially rescale our [Eu/Fe] distribution to match that of the broader \rp\ halo population to then obtain a more comprehensive NS mass distribution with our ADMC method.
Recent efforts by the RPA have resulted in such a large sample of metal-poor halo stars for which neutron-capture element abundances (i.e., Sr, Ba, and Eu) have been measured. This data set allows quantifying the \rp\ contribution and potential enhancement level for each star.

We select all RPA stars with definite measurements of Ba and Eu (i.e., no upper limits) with metallicities, [Ba/Eu], and [Eu/Fe] abundance ratios within the bounds of our initial 29-star sample.
By selecting stars with a constrained [Ba/Eu] ratio \citep[about $-0.7$, within observational uncertainties;][]{sneden2008}, we ensure that our sample contains stars that likely have \rp-only progenitor source(s) for their neutron-capture elements.
We select the same range of stars within the RPA sample with the above cutoffs to ensure a good match to our 29-star sample.

After these cutoffs, any contributions by, e.g., the \emph{s}-process or a limited \rp\ operating in SNe will be relatively small given the low-metallicity and the cut in [Ba/Eu] \citep[although even ${[{\rm Ba/Eu}]}<-0.4$ has the potential to include such contribution;][]{gull2020}. In fact, our key elements, Th and Dy are not significantly affected by any such other contributions so we regard any contamination(s) as insignificant for the purpose of the present study. 

The RPA sample of metal-poor stars contains 585 stars \citep[compiled from][]{hansen2018,sakari2018a,ezzeddine2020,holmbeck2020}.
Applying our cuts leaves 229 \rp\ stars within the abundance ratio ranges $-3.43 \leq [\mathrm{Fe/H}]\leq -2.06$, $-0.95 \leq [\mathrm{Ba/Eu}] \leq -0.40$, and $0.00 \leq [\mathrm{Eu/Fe}]\leq +1.80$.
The pared 229-star selection is a more statistically meaningful sample than our original sample of 29 stars with reported Zr, Dy, and Th measurements.
Therefore, we treat this 229-star sample as the current best and most accurate representation of the distribution of \rp\ material (i.e., [Eu/Fe]) in metal-poor halo stars.
For many of these stars, Th has not yet been measured.
The stars with available Th abundances have otherwise already been included in our MCMC sample (Table~\ref{tab:stars}).
With this larger \rp\ sample in hand, we next describe how we rescale our sample to achieve more realistic NS mass distributions from the ADMC results.

First, we bin the [Eu/Fe] abundances of the 229 RPA stars and our 29 stars in increments of 0.2\,dex and count how many stars are in each [Eu/Fe] bin.
This counting is to assess relative differences occurring from the biased selection of our sample.
We then use the bin counts to determine how much weight each ADMC-calculated result (DNS mass pair) should be assigned to produce a more unbiased, rescaled mass distribution.

To illustrate the necessity for such re-scaling, we provide the following example.
Our 29-star sample only contains two stars with $0.4\leq [{\rm Eu/Fe}] < 0.6$ (Stars 7 and 8 in Table~\ref{tab:stars}), whereas a plurality of metal-poor, \rp\ stars with [Eu/Fe] in the halo is found in this range \citep[see Figure~3 in][]{holmbeck2020}.
Therefore, we assign the mass results for Stars 7 and 8 a larger proportional weight to reflect that the halo contains a relatively larger fraction of these moderately enhanced \rp\ stars than what is suggested by our sample.
We note that this rescaling procedure can only be accomplished for mass \emph{distributions} (i.e., histograms) and cannot be applied to the individual result visualizations in Figure~\ref{fig:m1m2}.
Consequently, contributions by the \rii\ stars to the ADMC mass distribution are attenuated, while the contributions of our \ri\ and \ro\ stars are given more relative weight, reflective of a larger body of metal-poor, \rp-enhanced halo stars.

Figure~\ref{fig:nstar} shows the NS mass distribution for each EOS choice after applying this rescaling procedure.
These results are compared to parameterized fits to existing NS systems in the Milky Way.
The distribution labelled ``DNS" is the same as the prior given to the MCMC (a Gaussian with $\mu\approx 1.35$\,M$_\odot$ and $\sigma\approx 0.1$\,M$_\odot$), representing the handful of binary NS systems in the Galaxy today with precisely measured masses.
The millisecond pulsar (``MSP") distribution was taken from the fit presented in \citet{antoniadis2016}; these binaries are not exclusively NS-NS systems.

The results for many EOSs in Figure~\ref{fig:nstar} fall roughly within the DNS distribution, with some low-mass outliers.
Noticeably, the stiff EOS H4 cannot be reconciled with the present-day DNS distribution.
This discrepancy could provide additional, indirect support for ruling out H4 as a viable EOS choice.
Interestingly, no EOS choice reproduces the high-mass population of the MSP distribution.
This lack of high-mass NSs in our predicted DNS systems could provide another indication that NSs found in systems with non-NS companions have different evolutionary tracks than those with NS companions.

\section{Interpretations}
\label{sec:interpretations}

Taking these scaled results (and underlying assumptions) into account, our ADMC method can successfully reproduce the present-day Galactic distribution of DNS systems while simultaneously explaining the majority of heavy-element material in metal-poor \rp\ halo stars.
Supposing that all of our initial assumptions hold, the face-value interpretation of our results implies that NSMs could very well have been the progenitors of very metal-poor \rp-enhanced stars.
Furthermore, our results show that the DNS distribution in the Galaxy is roughly time-invariant since our reconstructed DNS systems agree with measurements of present-day ones.
However, there is no implicit need for this agreement to be the case.
Hence, we here reconsider some of our initial assumptions to analyze the underlying implications of this agreement.

\subsection{Stellar Sample Distribution}

	\begin{figure*}[t!]
	\centering
    \includegraphics[width=0.325\textwidth]{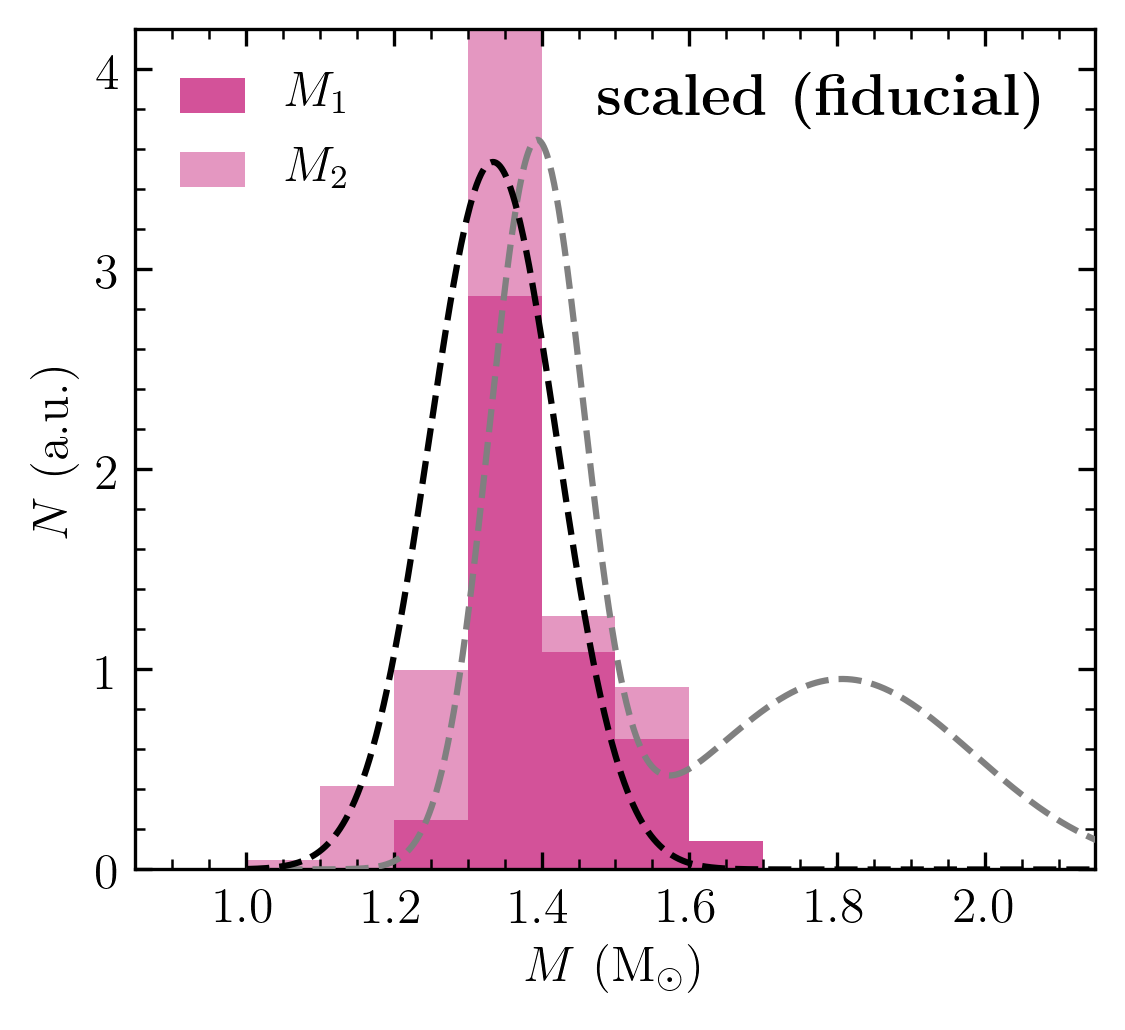}
    \includegraphics[width=0.325\textwidth]{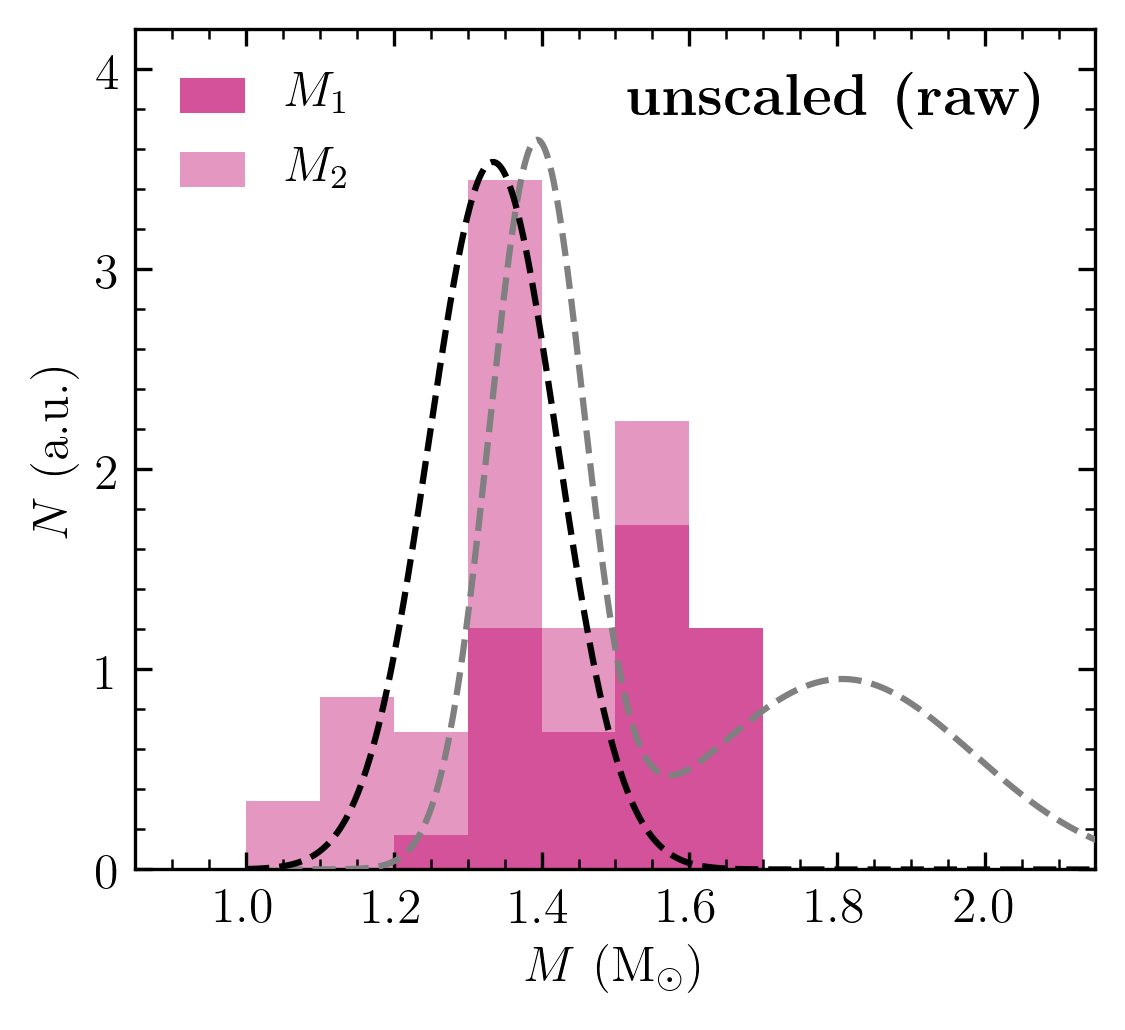}
    \includegraphics[width=0.325\textwidth]{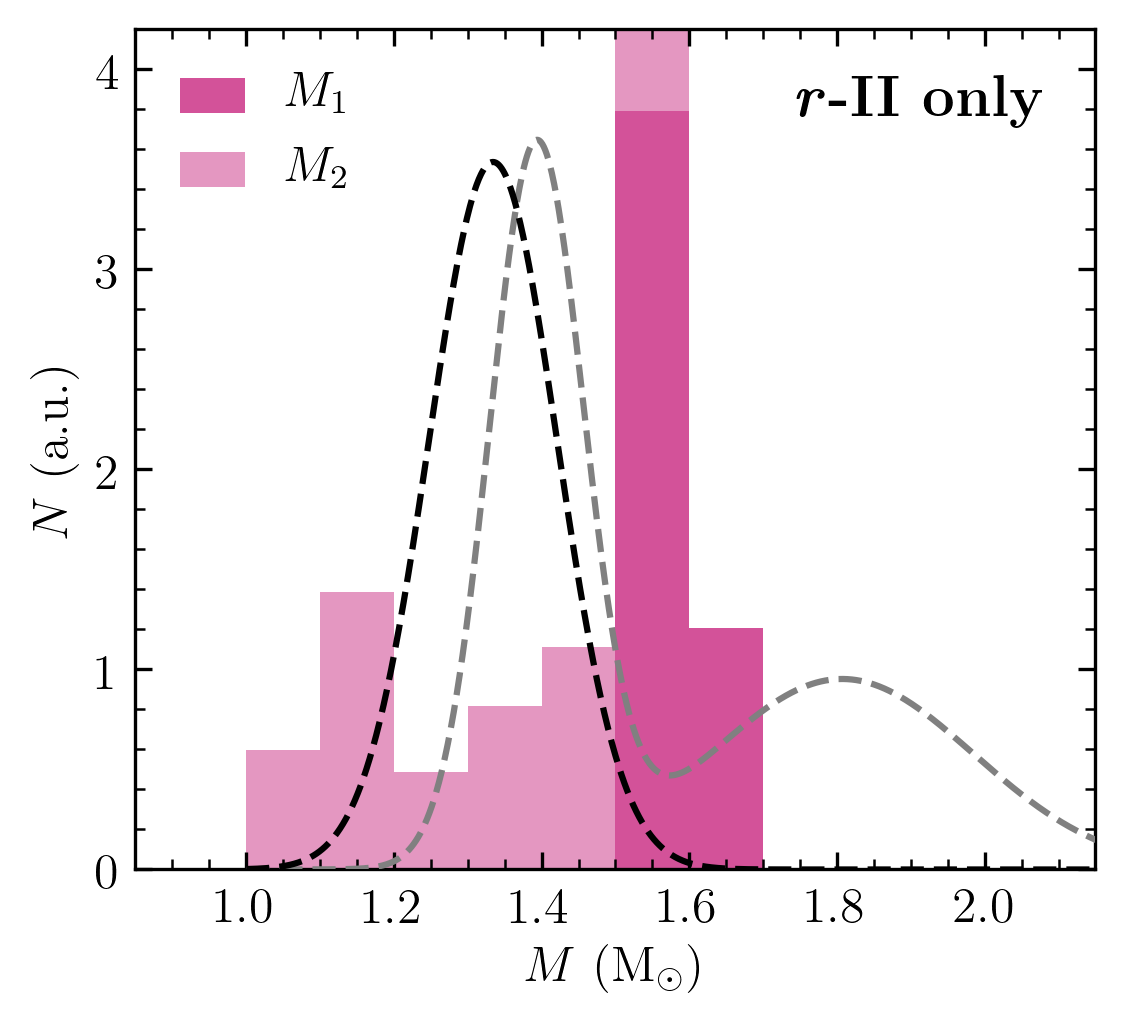}
    \caption{NS mass distributions for EOS DD2 with the scaling to metal-poor, \rp\ stars (left), when the scaling is removed (middle) and when only the \rii\ stars are considered (right). The black and gray dashed curves indicate the DNS and MSP mass distributions, respectively, as in Figure~\ref{fig:nstar}. Note that the left plot is the same as in Figure~\ref{fig:nstar}.\label{fig:dd2_norii}}
	\end{figure*}

Let us start by addressing the first assumption: that NSMs contribute the majority of the \rp\ element material in the Galaxy, especially the actinides. 
In the previous section, we have attempted to mitigate our initial selection bias by using a metal-poor star sample that spans a range of \rp\ enhancement levels.
However, what if other halo \rp\ stars that are not represented in our 29-star sample were to have different \rp\ progenitors?
To explore this possibility, we remove the applied scaling and take our the 29-star results at face value, i.e., we assume that these stars are true and wholly representative NSM descendants.

The left and middle panels of Figure~\ref{fig:dd2_norii} show the NS mass distribution for the DD2 EOS with and without the scaling procedure, respectively (results for the other EOSs are found in Supplemental Materials).
Overall, the ``raw" results (middle panel) do not well reproduce the mass distribution as set by the current Galactic NS population. Instead, for most EOSs, the results tend to favor the more asymmetric binary systems compared to what is presently observed. In fact, we find that multiple distributions from different EOSs are double-peaked rather than populating a single Gaussian distribution.
Taken at face value, our unscaled results thus seemingly imply that NSMs could only have been the progenitors of very metal-poor \rp-enhanced stars if those systems were mass-asymmetric. Consequently, if NSMs are primarily mass-\emph{symmetric} (as suggested by present day observations), then such NSMs could not have been the \rp\ progenitors of our metal-poor stars with extreme \rp\ enrichments.

We note that this conclusion could readily arise from either an observational bias associated with our sample selection, or have an underlying physical reason.
Observationally, Th in metal-poor stars is more straightforward to identify in high-resolution spectra when the star exhibits higher \rp\ element abundances. Hence, a natural bias exists in our sample towards detected Th spectral lines (and hence a strong prevalence towards higher abundances).
Although many upper limits on Th are available for additional stars, we do not use them in the ADMC procedure.
As a result, if our 29 stars thus inherently represent an extreme subset with only sufficiently high measured Th abundances, then it naturally follows that we would predict corresponding extremes on the DNS masses (following the matching of those high Th abundances). If correct, symmetric NSMs would have a tendency to produce less Th per unit of gas to dilute their yields into. This would lead to lower \rp\ levels found in metal-poor stars which would then have the tendency to lead to weaker Th lines that are harder to detect and measure, especially at low metallicities. 
In part, this limitation and associated bias in our sample is what the original scaling procedure attempts to mitigate.

At face value, this exercise suggests that many NS binary systems in the early universe (at least those that produced measurable levels of Th) could have been asymmetric, which makes them distinctly different from present-day binary NS systems. 
More generally, these results imply that low-$q$ systems were perhaps more common at low metallicities. Or early star forming regions hosted multiple binary systems of which all but the $q\approx 1$ systems evolved and inspiraled on faster timescales. Then, only those $q\approx 1$ systems survived to the present day due to significantly longer delay times.

\subsection{Different \rii\ progenitors}
\label{sec:rii_only}

Without the scaling procedure, our results cannot simultaneously account for the \rp\ material present in only the stars in Table~\ref{tab:stars} while at the same time reproducing the present-day DNS mass distribution.
Specifically, we have already commented on how the \rii\ stars tend to suggest very asymmetric NSMs as their \rp\ progenitors.
Since \rii\ stars have very clean abundance patterns (i.e., free from or only minimally contaminated by the \emph{s}-process), we now focus on interpreting our results when only considering the \rii\ stars.
We explore the ramifications of NSMs as dominant \rp\ sites if only the \rii\ stars are considered as true NSM descendants.

The right panel of Figure~\ref{fig:dd2_norii} shows the NS mass distribution derived when only considering the \rii\ stars present in our sample. (As this subset is also not representative of identified halo \rii\ stars, we scaled the sub-sample accordingly.)
If we assume that the \rii\ stars retain the cleanest \rp\ signatures---and are therefore the best tracers of individual NSM events---then our results imply a dominance of asymmetric systems in the early universe ($[{\rm Fe/H}]\lesssim -2$). 
The \rii-star-only derived DNS mass distributions suggest a preference for a large primary mass with $M_1\approx 1.6\,{\rm M}_\odot$ and a lower secondary mass of $M_2\approx 1.1\,{\rm M}_\odot$.
These mass distributions significantly diverge from our fiducial results that were based on constructing a representative sample of metal-poor stars with various \rp\ enhancement levels (Figure~\ref{fig:nstar}). However, it is important to acknowledge that there is no physical reason that our resulting mass distribution \emph{must} agree with those of present DNSs in the Milky Way.
Keeping this in mind, we further explore what can be learned if this extreme asymmetry were to be correct, at least for early universe DNSs. Specifically, we discuss implications regarding the existence of asymmetric DNSs within metal-poor environments that likely hosted \rii\ stars. 

The \rp-stars in the Galactic halo are believed to have originated in dwarf galaxies that were later accreted by the Milky Way, as evidenced, e.g., by the presence of \rii\ stars in the ultra-faint dwarf galaxy (UFD) Reticulum~II \citep[Ret II;][]{ji2016b}, \ri\ stars in Tucana~III \citep[Tuc III;][]{hansen2017}, and dynamical groups identified by \citet{roederer2018} and \citet{yuan2020}.
In this context, it is interesting to consider these dwarf galaxy host systems to learn more about the environments in which the progenitor NSs must have formed. There are currently two basic scenarios that attempt to explain the origin of \rp\ stars with different enhancement levels. The first scenario is based on the observational findings regarding Ret II and Tuc III; the highly enhanced \rii\ stars likely formed in smaller dwarf galaxies (i.e., Ret II analogs) where any \rp\ yields were diluted into less gas than what could occur in an increasingly more massive galaxy (such as Tuc III). The latter would then host the more diluted \ri\ stellar signature. The second scenario is based on recent hydrodynamical modeling of small galaxies that experienced an \rp\ event. \citet{tarumi2020} find that \rii\ stars form after such an event occurs in the center of the host galaxy, and \ri\ stars form as a result of an \rp\ event having occurred at the virial radius of the galaxy. This possibility is also discussed in \citet{safarzadeh2019}, which focuses on \rp\ enrichment by fast-merging, high-eccentricity DNSs within the virial radius of UFDs.

When interpreting our results regarding the progenitor nature of \rii\ stars in our sample, both these two scenarios lead to challenges that will eventually require resolution; i.e., are high-asymmetry systems more likely to occur and merge in UFDs like Ret II? For now, we simply present these challenges to highlight that more work is needed to understand the nature and progenitors of \rii\ stars, but more importantly, that insights into the environment of \rii\ star formation is a critical ingredient to carefully characterize \rp\ sites and ensure self-consistent progress. But while likely not straightforward to evaluate, it has become clear that host environment must be simultaneously taken into consideration.

With regard to just the \rii\ stars, the DNS progenitors appear to have been 
significantly asymmetric systems. 
In the context of the first possible scenario---that \rii\ stars form in Ret II-like UFDs---our results suggest that asymmetric NS binary systems are restricted to occur in the smallest dwarf galaxies (and \ri\ stars in more massive, Tuc III-like systems).
According to the second scenario, the results imply that asymmetric systems are restricted to merging only in or near the centers of (small) dwarf galaxies, and that (nearly) symmetric NSMs occur in the outskirts of their host galaxy.

Both scenarios bear significant challenges within our model. 
How do we reconcile our apparent correlation of UFD mass with DNS asymmetry?
The asymmetry could be explained if these \rp-producing DNSs were the remnants of massive Population~III (metal-free) binaries.
Alternatively, it is possible that asymmetric systems experience higher natal kicks, and therefore the DNSs that survive to merger quickly would occur closer to the centers of their host galaxies (as opposed to being ejected out of the galaxy were they near the virial radius).
Either way, any DNS present in such a small, early galaxy would have needed to have a short delay time, of order 100--1000~Myr in order to plausibly provide conditions for \rii\ stars to form \citep{bland-hawthorn2015,ji2016b,safarzadeh2019}.

\begin{figure*}[t]
	\centering
    \includegraphics[width=0.85\columnwidth]{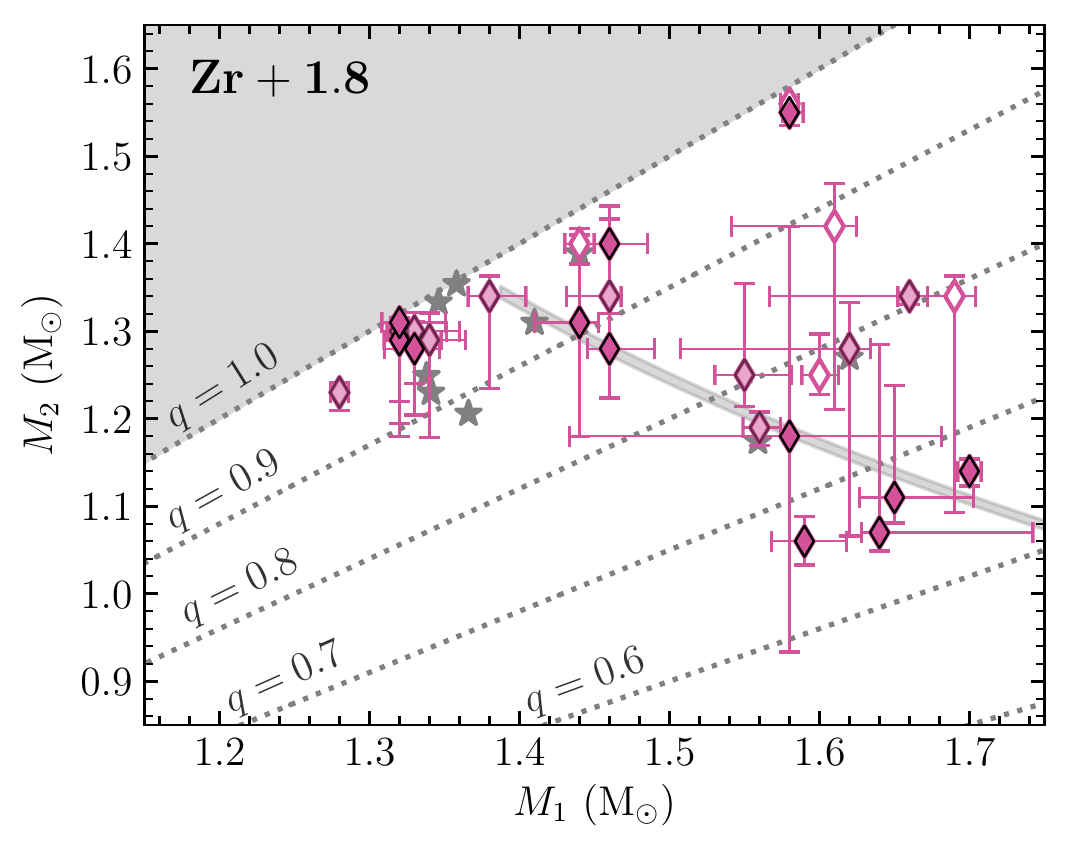}\qquad\quad
    \includegraphics[width=0.85\columnwidth]{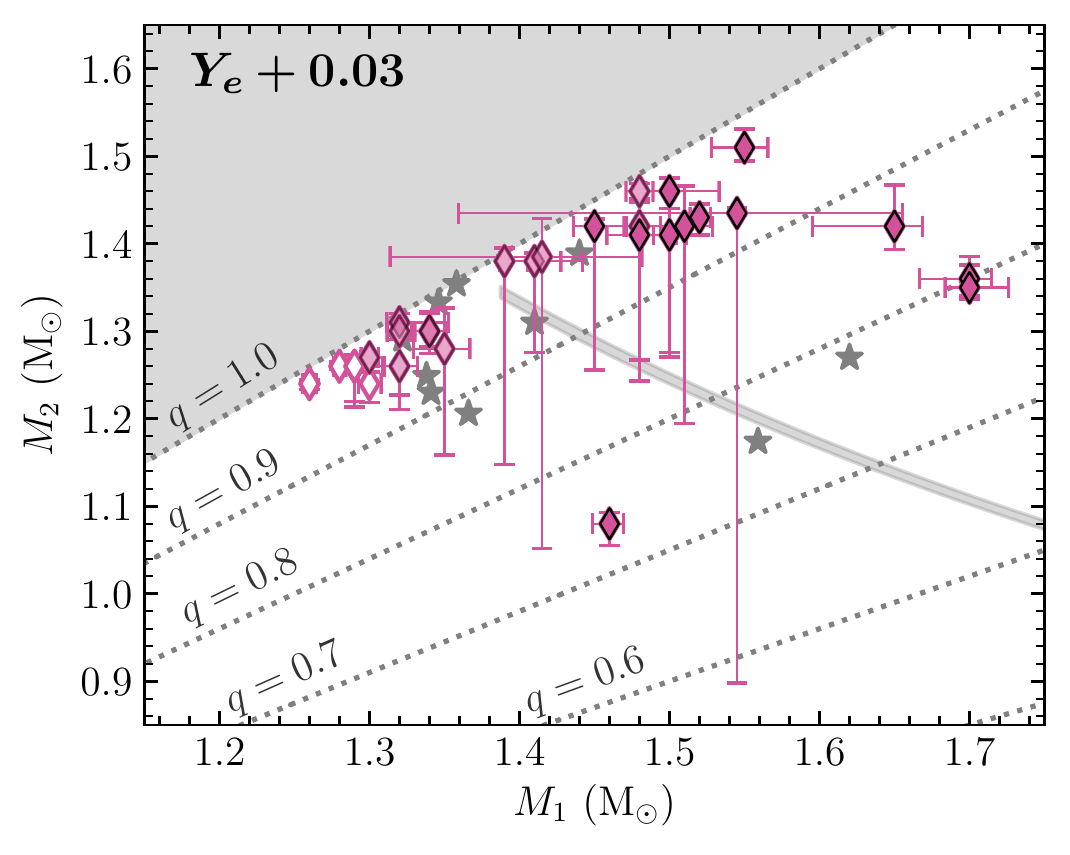}
    \caption{Mass solutions for EOS choice DD2 when the dynamical component contributes some Zr abundance to the total outflow mass (left) and when the composition of the disk is increased by $Y_e+0.03$ (right).\label{fig:dd2_zr18_p03}}
	\end{figure*}

We note, though, that a simple solution may also be possible: that there are no such puzzling correlations. Instead, perhaps asymmetric DNS mergers were simply prevalent in the early universe and thus are the only progenitors of \rp\ material observed in metal-poor halo stars. In that case, the lower observed \rp\ enhancement levels (\ri\ and \ro\ stars) may have arisen from dilution of the \rp\ yields at a later time, e.g., following the accretion of another smaller galaxy, before any \rp\ stars would have formed. 
Galactic chemical evolution and population synthesis studies will thus help to further explore the role of DNS with mass asymmetry and how they might have contributed to the enrichment of \rp\ elements within their host galaxies. Such studies should also include to what extent other \rp\ sites, e.g., collapsars \citep{Surman:2005kf,siegel2018} or other exotic supernovae \citep{winteler2012} are required to explain the observed signatures.

\section{Exploring Model Variations}
\label{sec:variations}

After discussing how asymmetric systems appear to be responsible for producing \rii\ stellar signatures, we now focus on how the predicted frequency of these extreme systems would change based on input variations used in our model. 
Following the assumptions outlined in Section~\ref{sec:results}, we consider what happens if the composition of the dynamical ejecta and disk wind is less (or more) neutron-rich, and if our results are robust under a uniform prior.
First, we address the composition. Modern calculations take increasing care to include neutrino effects on the composition of astrophysical ejecta. However, there always remains a chance that future simulations might predict higher (or lower) $Y_e$ values for NSM outflows.
We explore such variations in this section by artificially altering the ejecta composition in each component.

\subsection{Zr Contribution by the Dynamical Ejecta}

We start with the dynamical ejecta. What if this material contributes some significant amount of Zr to the final abundances?
As a straightforward case, we artificially increase the final Zr abundance of the dynamical ejecta by 1.8~dex, which increases the $\log\epsilon(\rm{Zr/Dy})$ abundance from about $-2.05$ to $-0.25$.
This addition mimics a very weak \limr\-process contribution which could potentially be achieved in dynamical outflows \citep[see, e.g.,][]{radice2018b}.
We rerun the ADMC method with this adjustment to the dynamical ejecta abundances and present the new solutions for DD2 in the left panel of Figure~\ref{fig:dd2_zr18_p03}.
(Solutions for the other EOS as well as the scaled NS mass distributions are included in Supplemental Materials.)

After this abundance increase, many of the solutions for \rii\ stars across the six EOSs suggest more symmetric DNS systems.
These solutions are effectively the opposite of our previous results, i.e., that the highly enhanced \rp\ stars generally require asymmetric NSMs. 
With the dynamical component contributing a larger amount of Zr to the total ejected Zr, there is less dynamical mass necessary to obtain the same ejected Zr mass as before.
Decreasing the dynamical mass can be achieved by either increasing the total mass of the DNS system or by increasing $q$ towards more symmetric values.
We find many of the individual solutions for the \rii\ stars prefer $q\sim 1$ as a way of decreasing the overall Zr contribution by the dynamical component, explaining why some solutions for the \rii\ stars become symmetric.

\subsection{Neutron-Richness of the Wind Outflows}

Next, we simulate the effect of added neutrino interactions in the disk wind outflows by adjusting the relative contribution of each tracer based on their initial $Y_e$.
We re-weigh the contribution by each tracer, effectively increasing the $Y_e$ of the disk outflows by 0.03 (roughly 10\%).
The abundances of main \rp\ elements, especially the actinides, are only sensitive to variations in the low-$Y_e$ ejecta.
Therefore, the final Zr abundance is nearly unaffected when the $Y_e$ is slightly increased.
On the other hand, the simulated Dy and Th abundances show an appreciable decrease when the $Y_e$ is increased even slightly, especially for merger remnants that survive collapse for a long time ($\tau = \infty$).
After slightly altering the nucleosynthesis yields from the disk outflows, we run ADMC again to see if different solutions are found.
We keep the same yields for the dynamical ejecta as in our initial results (with no increase to Zr).

The right panel of Figure~\ref{fig:dd2_zr18_p03} shows the ADMC results after a slight increase to the disk $Y_e$.
For many EOSs, the $Y_e$ increase produces \textit{more symmetric} systems with stronger peaks in the scaled mass distributions.
It may seem counter-intuitive that an increase in $Y_e$ of the disk wind would lead to more symmetric cases, but a careful interpretation of particularly Figure~\ref{fig:mdyn} can elucidate the cause of this apparent contradiction.
A binary NS system with a primary mass of 1.4\,M$_\odot$ will produce more dynamical ejecta at low $q$ than a system with a 1.2-M$_\odot$ primary.
Not only do these higher-mass systems provide enough dynamical ejecta to produce all the Th and Dy missing from the now relatively neutron-poor disk but they also produce NSs with masses that are more consistent with observations of existing DNS systems.
In addition, higher-mass systems are predicted to form smaller-mass disks around the massive remnant, and thus eject less Zr-rich material (see Figure~\ref{fig:mdisk}).
Therefore, the increase in Zr due to a slightly increased $Y_e$ is mitigated by preferring higher-mass systems that contribute less Zr in general from the disk wind component.
The increased $Y_e$ also leads to an effective decrease of the total Zr contribution from the disk outflows, allowing a less massive dynamical component to make up for the loss of light-\emph{r} production by the disk.

Overall, we see dramatic effects on the ADMC results when varying the ejecta composition. This indicates an area that would benefit from further improvements of NSMs compositions of the dynamically ejected mass and the wind outflow mass (potentially from kilonova observations) to ensure increasingly robust results when interpreting stellar abundances signatures. 

\subsection{Uniform Prior}
\label{sec:noprior}

The usefulness of Bayesian inference and prior knowledge is that expectations or physical limitations can be imposed on the otherwise reality-agnostic Monte-Carlo algorithm.
However, priors may restrict MCMC walkers from exploring a region of parameter space that may find a statistically more likely solution.
In this section, we relax the fourth assumption and apply a uniform prior (rather than a Gaussian one) to test if other, more or equally likely solutions are found in the parameter space.

The complete results of the MCMC runs with a uniform prior are shown in Supplemental Materials.
Figure~\ref{fig:m1m2_noprior} shows the $M_1$ and $M_2$ NS masses using DD2 and is duplicated here as an example.
While the abundance ratios of many of the stars still imply similar DNS systems as using a Gaussian prior, using a uniform prior sometimes yields very high-mass results.
A handful of input finds solutions around $M_{1,2} \approx 1.8$\,M$_\odot$.
When both $M_1$ and $M_2$ are at these high masses, two effects occur according to Equations~\ref{eqn:disk}--\ref{eqn:tau}.
First, at very high total masses, the merger remnant collapses immediately into a black hole, i.e., $\tau=0$\,ms.
In such prompt-collapse cases, there is theoretically no time for neutrino irradiation by the disk to change the composition of the ejecta, so the disk outflows stay extremely neutron-rich.
Secondly, the high NS masses prevent dynamical ejecta from escaping the merger.
Since the disk wind can now be neutron-rich enough to produce all three regions of interest in the abundance pattern---the \limr\ elements, the main \rp, and the actinides---no dynamical ejecta is required at all.
This particular solution only occurs for stars with sufficiently low Th/Dy (i.e., non-actinide-boost stars) such that the neutron-rich disk wind can synthesize all Th material required to match the input stellar abundance.

	\begin{figure}[t]
	\centering
    \includegraphics[width=0.9\columnwidth]{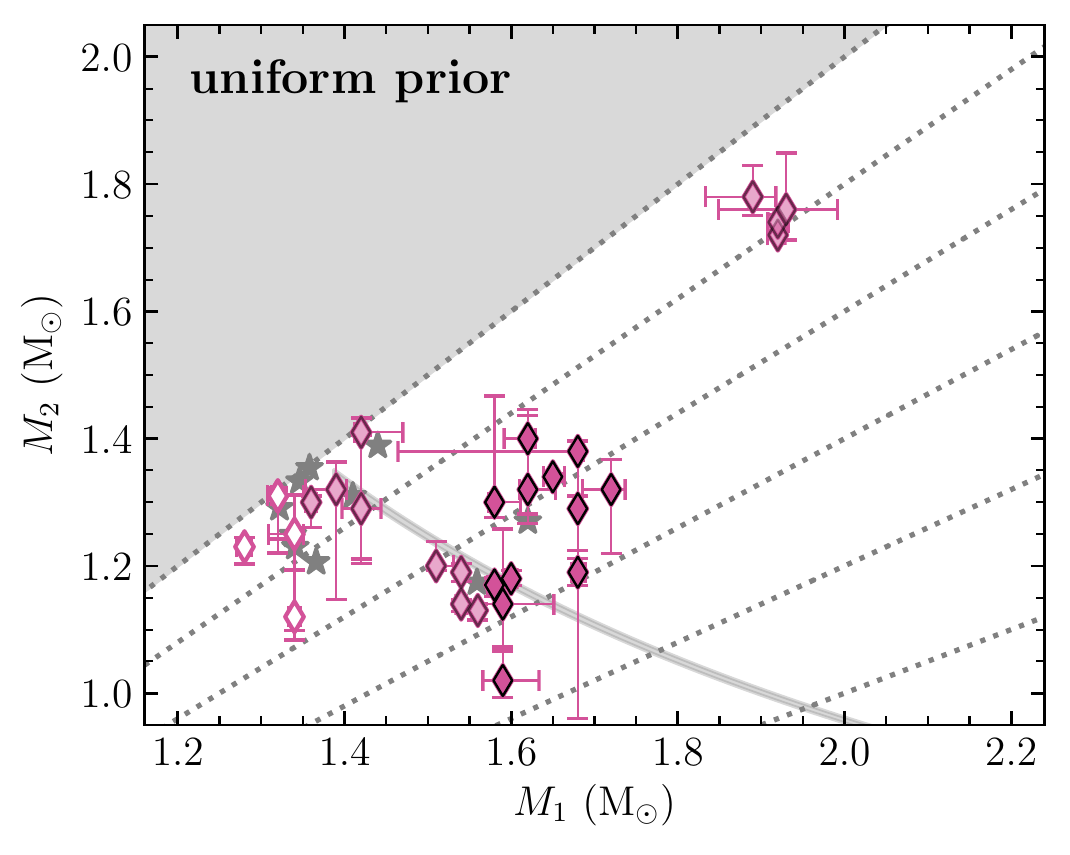}
    \caption[]{Mass results for DD2 when using a uniform prior. Note the axes are on a different scale from Figure~\ref{fig:m1m2}.\label{fig:m1m2_noprior}}
	\end{figure}

While these high-mass solutions are statistically as likely as the solutions using a Gaussian prior, the scenario for the production of the \rp\ elements is very different. 
At very high NS masses ($M_{1,2}\gtrsim 1.8$~M$_\odot$) the mass of the disk---and consequently the amount of disk wind ejecta---reaches a constant minimum of $10^{-3}$~M$_\odot$ as predicted by Equation~\ref{eqn:disk} (see Figure~\ref{fig:mdisk}).
Similarly, the dynamically ejected mass parameterized by Equation~\ref{eqn:dynk} also reaches a minimum as it plummets to 0~M$_\odot$ at high values of $M_{\rm tot}$ (see the right panel of Figure~\ref{fig:mdyn}).
Therefore, every more massive $M_1$-$M_2$ combination is predicted to produce the exact same values for $m_{\rm wind}$ and $m_{\rm dyn}=0$ (as well as the same abundance patterns since a large $M_{\rm tot}$ leads to prompt-collapse remnants).
This high-mass corner of the parameter space represents, for certain stellar inputs, a likelihood plateau where all solutions are similarly successful and non-unique.
Within this current model, we cannot break the degeneracy between non-unique solutions.
Further investigation is needed to determine if these disk-only cases are compatible with other observational and theoretical evidence not incorporated in our method.
For example, by comparing electromagnetic signals associated with GW events---or, equally importantly, the lack thereof---to determine if a disk-only case sufficiently produces not only the lanthanides, but also enough actinide material to account for the full range of \rp\ elements in metal-poor stars.

We have explored different variations on both the implicit and explicit assumptions in our model and find that the outflow composition indicates the largest area for improvement. 
In this context, the variations on ejecta composition reveal that the ratio of weak to main \rp\ production---or Zr/Dy, as we represent that ratio---is absolutely critical, and will require future confirmation.

There remains one model variation that we have not addressed which is the possibility of additional, subdominant sites that contribute significantly to the observed neutron-capture element pattern in \rp\ metal-poor stars.
For example, a second site might pollute star-forming gas with an abundance of Zr, and hence, a higher weak-to-main \rp\ ratio, than an NSM event enriching the same gas.
In this case, the particular NSM contributing the majority of main \rp\ material would require less Zr to be produced in its outflows.
As described in the interpretation of our initial results, restricting weak-to-main \rp\ ratios to lower values would likely push the MCMC solutions to higher mass asymmetries.
This important variation on the outflow ejecta can either have theoretical or astrophysical origins, such as if the ejecta is less (or more) neutron-rich than predicted by simulations, and/or if more than one site contributes to each stellar \rp\ signature.
These important questions will require further investigation that has the potential to strengthen this model and its results.

\section{Summary and Discussion}

This work introduces a novel and innovative technique in which stellar abundances work hand-in-hand with simulation data in order to uncover histories of past astrophysical phenomena in a way that has thus far not been attempted in literature.
We achieve this new connection by applying the best current fits to data obtained from hydrodynamical simulations of \rp\ nucleosynthesis describing the total wind and dynamical ejecta from NSMs as a function of the individual NS masses.
Next we use state-of-the-art nucleosynthesis network solvers to find the compositions---and therefore the total \rp\ yields---of these two ejecta components.
Then, we input observed stellar \rp\ abundances using the ratios of \limr-to-main-to-actinide elements and postulate that unique NSMs were responsible for the \emph{entire} observed \rp\ patterns in these stars.
Working backwards from the observed stellar abundance ratios, this method uses an MCMC algorithm to find the optimal masses of the two NSs that would have merged to produce those \rp\ abundances. We strive for self-consistency wherever possible, both in the nuclear data and in the EOS that sensitively shapes the NSM ejecta when combining current theoretical and observational results in this new way.

Our main results are summarized as follows:
\begin{itemize}
    \item If NSM are responsible for the majority of the main \rp\ material observed across \rp\ enhanced Galactic metal-poor halo stars, then those progenitor NSMs had masses similar to present-day DNSs in the Galaxy.
    \item There is a mass-asymmetry of the progenitor system that increasingly occurs with higher levels of \rp\ enhancement; the \rii\ stars require the largest asymmetries among progenitor systems, while stars displaying low \rp\ enhancements can be explained by mergers of somewhat equal-mass NSs.
    \item If the \rii\ stars bear the only pure NSM signatures, then their progenitor mass asymmetries either indicate that past merging DNS systems were distinctly different from present-day DNSs, or alternatively reveal previously unexplored correlations with DNS formation and host galaxy environment.
    \item Our results using EOSs with larger values of $M_{\rm TOV}$ predict slightly more symmetric binaries, indicating that the effect of EOS on the MCMC results manifests primarily from the maximum, non-rotating NS mass rather than the compactness ($\mathcal C$).
\end{itemize}

We explored some variations in the assumptions of our model to try to identify key sensitivities to our theoretical inputs. Of particular note is the electron fraction of the disk wind ejecta. If the disk wind is slightly more proton-rich than current simulations estimate, then more massive---and more symmetric---NS binary systems would best explain the \rp\ abundances of metal-poor stars. Thus developments in hydrodynamical simulations and neutrino transport will have bearing on our conclusions. Detailed treatment of neutrino flavor transformation in NSM ejecta is essential towards understanding the outflow composition and can reveal whether the disk wind can reach significant levels of neutron-richness or not.

We showed in Section~\ref{sec:noprior} that, when allowed to explore a wider parameter space, the MCMC walkers find degenerate solutions at high NS masses.
NSM simulations and ejecta mass characterizations from hydrodynamical simulation data at extreme primary masses and DNS mass-asymmetries may also explore the lower ejecta limit for NSMs and whether alternative, high-mass solutions are viable.
Currently available parameterizations of the dynamical and disk wind ejecta masses are insensitive to these high masses.

With more observational data likely to become available soon through dedicated efforts, e.g., the RPA collaboration, we will be able to straightforwardly adapt and update our inputs. Ideally, complete, self-consistent stellar abundances of the entire \rp\ pattern (from the \limr\ elements to the actinides) as well as measurements of second-peak elements (e.g., tellurium) could be used as additional constraints on the nature of the predicted DNS progenitor systems.
Although not explored in this paper, nuclear uncertainties that sensitively affect the production of the rare-earth and actinides elements could also impact our results.

We also look forward to complementary developments in population synthesis studies, which may be able to elucidate whether high mass-asymmetry DNS systems are more likely to form from low-metallicity gas, or if they can form sufficiently frequently for NSMs to be the primary producers of \rp\ material as observed in the metal-poor halo stars. 
Additionally, future LIGO/Virgo observations and any electromagnetic counterparts can offer further constraints on the likelihood that NSMs produced a majority of the \rp\ material in the Galaxy. 
On the extremely asymmetric side, neutron star-black hole mergers (NSBHs) may also eject neutron-rich material. We table the possibility of NSBH origins for the metal-poor stars in this work for future discussion.
Updated NSBH event rates from LIGO/Virgo could constrain how often NSBHs contribute as \rp\ sources and whether their (expectedly) low rates---but high ejecta yields---are comparable to the more frequent---but less prolific---NSMs.

This work demonstrates a unique and adaptable route by which the elemental signatures of metal-poor stars can place constraints on the first generation(s) of stars and stellar populations in the universe and, potentially, on the nuclear EOS. As more experimental, theoretical, and observational data are gathered, we look forward to improving our inputs, modernizing our method, and refining our results.


\acknowledgements

EMH thanks Will Newton, Matt Caplan, Benoit C\^{o}t\'{e}, and Chris Belczynski for the exciting and enlightening discussions about the EOS and Galactic chemical evolution.
This work was initiated by discussions at the ``\rp\ Sources in the Universe Workshop," which was supported by National Science Foundation under Grant No.\ PHY-1430152 (JINA Center for the Evolution of the Elements).
AF acknowledges support from NSF grants AST-1255160 and AST-1716251. 
GCM, RS, and TMS were partially supported by the Fission In R-process Elements (FIRE) topical collaboration in nuclear theory, funded by the U.S.\ Department of Energy (DoE).
DoE provided additional support through contract numbers DE-FG02-02ER41216 (GCM), DEFG02-95-ER40934 (RS and TMS) and DE-SC0018232 (SciDAC TEAMS collaboration; RS and TMS).
RS and GCM also acknowledge support by the National Science Foundation Hub (N3AS) Grant No.\ PHY-1630782. 
MRM was supported by Laboratory Directed Research and Development program under project number 20190021DR.
Los Alamos National Laboratory is operated by Triad National Security, LLC, for the National Nuclear Security Administration of the DoE (Contract No.\ 89233218CNA000001).
RF is supported by the Natural Sciences and Engineering Research Council (NSERC) of Canada through Discovery Grant RGPIN-2017-04286, and by the Faculty of Science at the University of Alberta.

\software{
\texttt{corner} \citep{corner}, \texttt{emcee} \citep{foreman2013}, \texttt{Matplotlib} \citep{hunter2007}, \texttt{NumPy} \citep{vanderwalt2011}, \texttt{SciPy} \citep{2020SciPy-NMeth}
}


\bibliography{main.bib}

\end{document}